\documentclass[a4paper,10pt]{article}
\usepackage[utf8]{inputenc}
\usepackage[a4paper, textwidth=15cm, top=2cm, bottom=2cm]{geometry}
\usepackage{authblk}
\usepackage{graphicx}
\usepackage{subfigure}
\usepackage{amssymb}
\usepackage{url}
\usepackage{color} 
\usepackage{caption}
\usepackage{xcolor}
\usepackage{overpic}
\usepackage{units}
\usepackage{booktabs}

\newcommand{\cenns}{CE$\nu$NS}
\newcommand{\nucleus}{NUCLEUS}
\newcommand{\alo}{Al$_2$O$_3$}
\newcommand{\cawo}{CaWO$_4$}
\newcommand{\rateunit}{counts/(keV$\cdot$kg$\cdot$day)}

\title{Exploring \cenns{} with \nucleus{} at the Chooz Nuclear Power Plant}

\begin{document}


\author[1]{G.~Angloher}
\author[2,3]{F.~Ardellier-Desages}

\author[1,4]{A.~Bento}
\author[1]{L.~Canonica}
\author[5]{A.~Erhart}
\author[1]{N.~Ferreiro}
\author[6]{M.~Friedl}
\author[6]{V.M.~Ghete}
\author[1]{D.~Hauff}
\author[6,7,*]{H.~Kluck}
\author[5,*]{A.~Langenk\"{a}mper}
\author[2,3]{T.~Lasserre}
\author[2]{D.~Lhuillier}
\author[5]{A.~Kinast}
\author[1]{M.~Mancuso}
\author[8]{J.~Molina Rubiales}
\author[5]{E.~Mondragon}
\author[8]{G.~Munch}
\author[2]{C.~Nones}
\author[5]{L.~Oberauer}
\author[2]{A.~Onillon}
\author[5]{T.~Ortmann}
\author[5]{L.~Pattavina}
\author[1]{F.~Petricca}
\author[5]{W.~Potzel}
\author[1]{F.~Pr\"{o}bst}
\author[6,7]{F.~Reindl}
\author[1,*]{J.~Rothe}
\author[6,7]{J.~Schieck}
\author[5]{S.~Sch\"{o}nert}
\author[6,7]{C.~Schwertner}
\author[2]{L.~Scola}
\author[1]{L.~Stodolsky}
\author[5]{R.~Strauss}
\author[2]{M.~Vivier}
\author[2,\footnote{Corresponding author},\footnote{Email: Victoria.Wagner@cea.fr}]{V.~Wagner}
\author[\space]{A.~Zolotarova\textsuperscript{2} \\(The \nucleus{} Collaboration)}

\affil[1]{Max-Planck-Institut für Physik, D-80805 München, Germany}
\affil[2]{IRFU, CEA, Universit\'{e} Paris Saclay, F-91191 Gif-sur-Yvette, France}
\affil[3]{APC, Astro Particule et Cosmologie, Universit\'e Paris Diderot, CNRS/IN2P3, CEA/Irfu, Observatoire
de Paris, Sorbonne Paris Cit\'e, 10, rue Alice Domon et Leonie Duquet, 75205 Paris Cedex 13, France}
\affil[4]{CIUC, Departamento de Fisica, Universidade de Coimbra, P3004 516 Coimbra, Portugal}
\affil[5]{Physik-Department, Technische Universität München, D-85748 Garching, Germany}
\affil[6]{Institut für Hochenergiephysik der Österreichischen Akademie der Wissenschaften, A-1050 Wien, Austria}
\affil[7]{Atominstitut, Technische Universit\"at Wien, A-1020 Wien, Austria}
\affil[8]{\'Electricit\'e de France, Centre nucl\'eaire de production d'\'electricit\'e de Chooz, Service Automatismes-Essais, 08600 Givet, France}

\maketitle
\begin{abstract}
Coherent elastic neutrino-nucleus scattering (\cenns{}) offers a unique way to study neutrino properties and to search for new physics beyond the Standard Model. 
Nuclear reactors are promising sources to explore this process at low energies since they deliver large fluxes of \newline (anti-)neutrinos with typical energies of a few MeV. 
In this paper, a new-generation experiment to study \cenns{} is described. 
The \nucleus{} experiment will use cryogenic detectors which feature an unprecedentedly low energy threshold and a time response fast enough to be operated in above-ground conditions. Both sensitivity to low-energy nuclear recoils and a high event rate tolerance are stringent requirements to measure \cenns{} of reactor antineutrinos. 
A new experimental site, denoted the Very-Near-Site (VNS), at the Chooz nuclear power plant in France is described. The VNS is located between the two 4.25\,GW$_{\mathrm{th}}$ reactor cores and matches the requirements of \nucleus{}. First results of on-site measurements of neutron and muon backgrounds, the expected dominant background contributions, are given. 
In this paper a preliminary experimental setup with dedicated active and passive background reduction techniques is presented. 
Furthermore, the feasibility to operate the \nucleus{} detectors in coincidence with an active muon-veto at shallow overburden 
is studied. 
The paper concludes with a sensitivity study pointing out the promising physics potential of \nucleus{} at the Chooz nuclear power plant. 
\end{abstract}


\section{Introduction}

The existence of neutral-current neutrino interactions implies the existence of elastic neutrino-nucleus scattering
\cite{freedmann74}.
The cross-section of this process \cite{Lindner:2016wff} is given in the Standard Model of Particle Physics (SM) by
\begin{equation}
 \frac{\mathrm{d}\sigma}{\mathrm{d}\mathrm{E}_{\mathrm{R}}}\,=\,\frac{\mathrm{G}^2_{\mathrm{F}}}{4\,\pi}\,\mathrm{Q}^2_{\mathrm{W}}\,\mathrm{F}^2(\mathrm{q}^2)\,\cdot\,\mathrm{m}_{\mathrm{N}}\, \left(1\,-\,\frac{\mathrm{E}_{\mathrm{R}}}{\mathrm{E}^{\mathrm{max}}_{\mathrm{R}}}\right)
 \label{eq:cenns_xsection}
\end{equation}
where $\mathrm{G}_{\mathrm{F}}$ is the Fermi constant, $\mathrm{Q}_{\mathrm{W}}$ is the nuclear weak charge, $\mathrm{m}_{\mathrm{N}}$ is the total mass of the nucleus,  $\mathrm{E}_{\mathrm{R}}$ is the nuclear-recoil energy, 
and $\mathrm{F}(\mathrm{q}^2)$ is the  nuclear form factor as a function of the momentum transfer q. The weak charge is given by $\mathrm{Q}_{\mathrm{W}}\,=\,\mathrm{N}\,-\,\mathrm{Z}\cdot(1-4\cdot\mathrm{sin}^2\theta_{\mathrm{W}})$, with N and Z being the number of neutrons and protons of the target nucleus, and $\theta_{\mathrm{W}}$ the Weinberg angle.
The maximum nuclear recoil energy $\mathrm{E}_{\mathrm{R}}$ is given by $\mathrm{E}^{\mathrm{max}}_{\mathrm{R}}\,=\,2\,\mathrm{E}_{\nu}^2/(\mathrm{m}_{\mathrm{N}}\,+\,2\,\mathrm{E}_{\nu})$, where $\mathrm{E}_{\nu}$ is the incident neutrino energy. 
For momentum transfers smaller than the inverse of the nuclear radius (typically for $\mathrm{E}_{\nu}\,\lesssim\,30$\,MeV) the form factor is close to unity and the scattering is coherent over all nucleons in a nucleus \cite{freedmann74}: this process is called coherent elastic neutrino nucleus scattering (\cenns{}). 
For larger momentum transfers the full coherence is no longer given and the form factor drops below unity.

\begin{figure}
 \centering
 \includegraphics[width=0.75\textwidth]{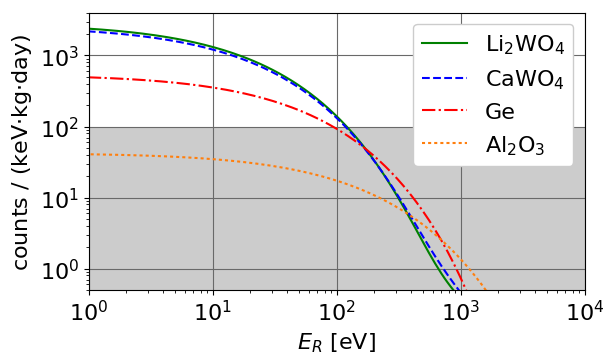}
 \caption{Differential \cenns{} count rate on Li$_2$WO$_4$ (solid green line), CaWO$_4$ (dashed blue line), germanium (dash-dotted red line) and Al$_2$O$_3$ (dotted orange line), calculated with the antineutrino flux expected at the Very-Near-Site 
 from both Chooz-B reactor cores. The reactor neutrino flux model follows~\cite{Tengblad:1989db} as parameterized in~\cite{guetleinPhD}.
 The \nucleus{} experiment aims to reach a background count rate of of 100\,\rateunit{}, indicated by the gray band. 
 }
 \label{fig:RatePlot}
\end{figure}

The coherent cross-section is boosted by a factor of N$^2$, and can exceed the cross-section of the standard neutrino detection methods, such as inverse beta-decay (IBD), by more than two orders of magnitude \cite{Akimov:2017ade}. 
Contrary to the enhancement of the cross-section, the experimental signature, i.e. the nuclear recoil, is suppressed in energy by the nucleus mass, $\mathrm{m}_{\mathrm{N}}$. 

Figure \ref{fig:RatePlot} shows the differential count rate of coherently scattered antineutrinos in different targets. 
For $\mathrm{E}_{\mathrm{R}}\,\leq\,100$\,eV, the \cenns{} count rate in a CaWO$_4$ or Li$_2$WO$_4$ crystal exceeds the rate on lighter nuclei such as present in Al$_2$O$_3$ by more than an order of magnitude. 
However, as the \cenns{} count rate decreases rapidly with increasing energy, an energy threshold well below 100\,eV is necessary to explore the \cenns{} signal with heavy targets. 

While for IBD the kinetic energy of the antineutrino needs to be at least 1.8\,MeV, \cenns{} does not have an energy threshold. Thus, \cenns{} provides a unique probe of the Standard Model (SM) at low energies: e.g. measuring the Weinberg angle, $\theta_{\mathrm{W}}$, at low momentum transfer \cite{Drukier:1983gj} or exploring fundamental neutrino properties such as the existence of a neutrino magnetic dipole moment \cite{Vogel:1989iv}. 
Moreover, as a neutral current interaction, \cenns{} is flavor insensitive, and, thus, a new probe to search for sterile neutrinos \cite{Kosmas:2017zbh}. 
Any deviation from the SM prediction may reveal new physics beyond the SM, such as non-standard neutrino interactions, i.e. modified V-A quark-neutrino couplings, or new exotic neutral currents, see e.g. \cite{Lindner:2016wff, PhysRevD.73.033005}. 
\cenns{} offers a broad spectrum of possible applications in nuclear physics \cite{Cadeddu:2017etk, PhysRevC.86.024612} and supernovae detection  \cite{Drukier:1983gj, PhysRevD.68.023005}. 
The enhancement of the \cenns{} cross-section allows for a miniaturization of neutrino detectors, from the typical tonne-size to kilogram- or even gram-scale in the case of \nucleus{}, and thus, a possible practical application in nuclear reactor monitoring \cite{Strauss:2017cuu}.
Furthermore, \cenns{} of solar and atmospheric neutrinos will become an irreducible background for future dark matter experiments searching for weakly interacting massive particles, which thus profit from an independent measurement of the cross-section. 

The first observation of \cenns{} was reported by the COHERENT collaboration at a $6.7\sigma$ confidence level \cite{Akimov:2017ade}. The COHERENT detector \cite{Akimov:2015nza}, operating at the Spallation Neutron Source (SNS) in a neutrino flux of $4.3\cdot10^{7}\nu/(\mathrm{s}\,\cdot\,\mathrm{cm}^2 )$, uses a 14.6 kg CsI[Na] scintillating crystal target. In contrast to reactor neutrinos, the neutrino beam produced at the SNS features a well defined energy spectrum reaching up to 50\,MeV, i.e. partially above the coherence regime of elastic neutrino-nucleus scattering.

\vspace{0.5cm}
This paper describes the \nucleus{} experiment at the Chooz Very-Near-Site (VNS), designed to study \cenns{} using reactor antineutrinos.
The VNS is presented in Section \ref{sec:VNS}, together with on-site muon and neutron attenuation measurements. 
Section \ref{sec:NuCleus} focuses on the concept of the \nucleus{} experiment at the VNS, whereas the \nucleus{} target detectors are described in details in Reference \cite{Strauss:2017cuu}. 
Section \ref{sec:DeadTime} demonstrates that the \nucleus{} detectors can be operated at the VNS with a moderate dead time induced by the muon-veto. 
The potential to measure the \cenns{} process with the \nucleus{} experiment at the VNS is presented in Section \ref{sec:Sensitivity}.


\section{The Very-Near-Site at the Chooz Nuclear Power Plant}
\label{sec:VNS}
\begin{figure}
 \centering
 \includegraphics[width=0.75\textwidth]{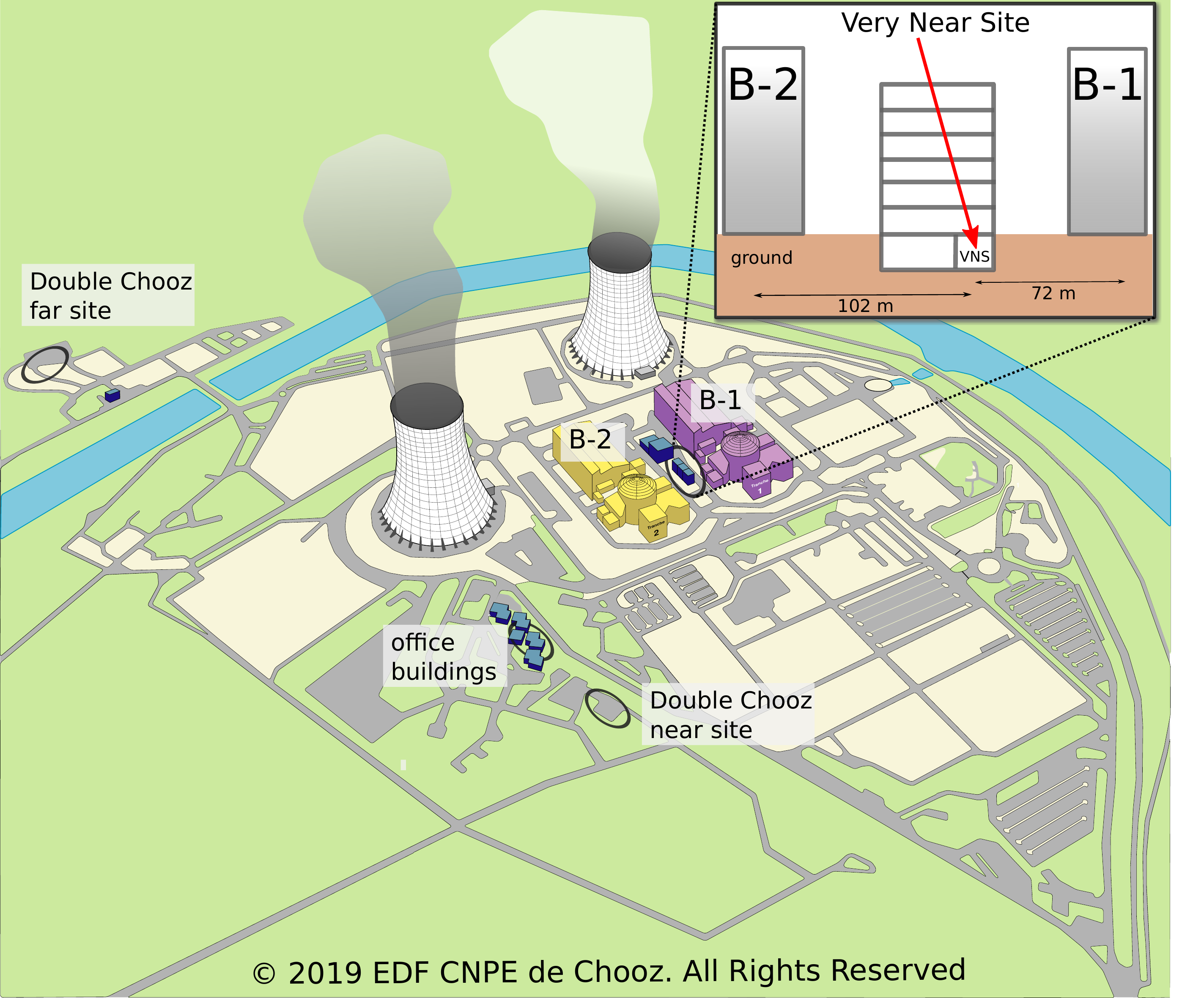}
 \caption{Sketch of the Chooz nuclear power plant and of the Very-Near-Site (VNS). The VNS is located in-between the two reactor buildings B-1 and B-2. Since 2008, the power plant hosts the Double Chooz experiment with two experimental sites, denoted the near and the far site. Office buildings close to the Double Chooz sites can be used in the future. 
The inset on the top right shows the office building between the reactor cores hosting the VNS. 
Figure credit to EDF CNPE de Chooz. }
 \label{fig:Chooz}
\end{figure}
The VNS is a new experimental site between the two power reactors of the Chooz-B plant, planned to host the \nucleus{} experiment. 
The nuclear power plant, shown in Figure \ref{fig:Chooz}, is operated by the French company Electricit\'e de France (EDF). 
The two N4-type pressurized water reactors, hereafter identified as B-1 and B-2, are separated by 160\,m, and their respective cores are located about 7\,m above the Chooz ground level. 
Each reactor runs at a nominal thermal power of 4.25\,GW$_{\mathrm{th}}$  and features a high duty cycle. The two cores are switched off for refuelling approximately one month per year during alternating periods. 

Nuclear reactors are one of the strongest artificial continuous neutrino sources. 
Reactor-$\bar{\nu_{e}}$ are produced via the beta-decays of the fission products of $^{235}$U, $^{238}$U, $^{239}$Pu and $^{241}$Pu with energies up to 10\,MeV. The mean $\bar{\nu_{e}}$ energy is around 1.5\,MeV \cite{Kopeikin:2012zz}.
Thus, reactor-$\bar{\nu_{e}}$ are expected to be in the fully coherent region for typical nuclear target elements, maximizing the \cenns{} cross-section. 
Assuming an average number of six $\bar{\nu_e}$ per fission with an average energy release of 200\,MeV, about $8\cdot10^{20}\,\bar{\nu_e}/\mathrm{s}$ are produced by a  4.25\,GW$_{\mathrm{th}}$ power reactor \cite{Kopeikin2004}.

Until 1998, the Chooz power plant hosted the long baseline ($\sim1000\,\mathrm{m}$) neutrino oscillation experiment CHOOZ \cite{Apollonio:2002gd}. The main result was an upper limit on neutrino oscillation in the $\bar{\nu_{e}}$ disappearance mode.  
Since 2008 the Chooz laboratory site is used by the neutrino oscillation experiment Double Chooz to measure the neutrino mixing angle $\theta_{13}$ \cite{ChoozNS}.

Thanks to these previous neutrino experiments, the Chooz reactor site has an existing infrastructure for scientists, including office, meeting and storage rooms, which can be made available for future projects. 
The acquired expertise and relationship developed in past collaborations with EDF greatly helps to establish the new site for a future neutrino experiment.

\subsection{Description of the VNS}
\label{sec:DescritionVNS}
The VNS is a 24\,m$^2$ room situated in the basement of a five-story office building at the Chooz power plant (see inset of Figure \ref{fig:Chooz}).
The small size of the room requires that the full setup does not exceed a volume of several cubic meters and a weight of the order of $\mathcal{O}(10\,\mathrm{t})$.
To support loads on the floor of $\mathcal{O}(5\,\mathrm{t}/\mathrm{m}^2)$, as expected for the  \nucleus{} shielding, a weight support platform will be installed.
Furthermore, minor modifications of the VNS are foreseen to meet the safety regulations.

Mechanical vibrations are a known cause of disturbance in cryogenic detectors. Vibrations can arise within the cryostat necessary to operate the detectors. In addition, vibrations may be introduced to the setup at the VNS by the turbines and generators of the nuclear power plant in the close vicinity.   
Further external sources of vibrations can be of seismic origin, as well as introduced by the elevator system, or movements in the office building.
A vibration measurement campaign is planned for 2019. 
Based on the campaign results, a system to attenuate the vibrations will be designed.

The baseline of the VNS to the two reactor cores is 72\,m to B-1, and 102\,m to B-2 respectively.
The expected neutrino flux at the VNS is about  $10^{12}\,\bar{\nu_{\mathrm{e}}}/(\mathrm{s}\cdot\mathrm{cm}^{2})$. 
At this distance, no neutron background from the reactor cores is expected (see Section \ref{sec:background}). However, for an  expected overburden of less than 10 meters of water equivalent (m.w.e.), see section below, special care of the cosmic-ray induced background needs to be taken. Therefore, a campaign to characterize the neutron- and muon-background at the VNS was performed from October 2017 until May 2018. 
The campaign results are presented below and used to estimate the upper limit on the trigger rate of the \nucleus{} muon-veto in Section \ref{sec:DeadTime}. Furthermore, they will be used to optimize the design of the compact shielding and to estimate the expected background count rate in the \cenns{} target detectors in the future.

\subsection{Measurement of the Muon Flux Attenuation}
\label{sec:MuonAttenuation}
\begin{figure}
 \centering
 \includegraphics[width=0.6\textwidth]{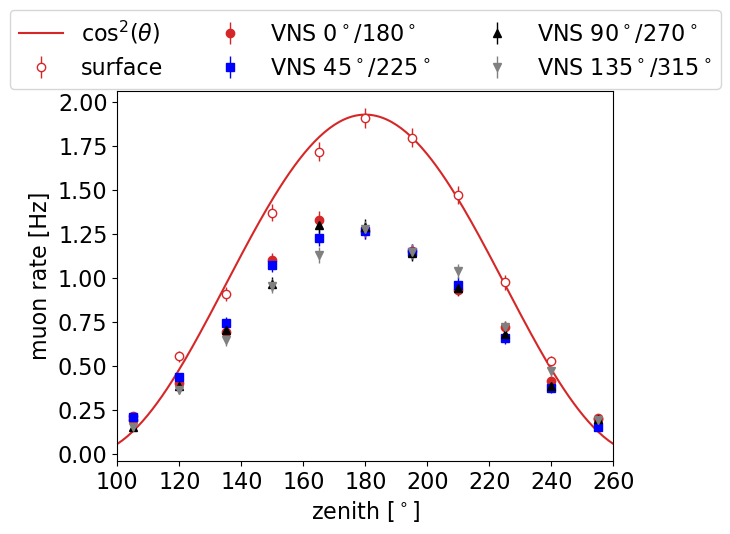}
 \caption[]{Measurement of the muon rate at the surface (open data points) and at the VNS (filled data points) as a function of zenith angle for different azimuthal orientations. The muon flux at the surface follows the expected cos$^2\theta$-law (solid red line). Uncertainties are statistical only.}
 \label{fig:muonMeasurement}
\end{figure}

The muon rate at the location of the VNS was measured with the \textit{cosmic wheel} developed by the Centre de physique des particules de Marseille (CPPM) for the ``Science \`a l'\'ecole'' outreach program~\cite{ruoeCosmique}.
The device consists of three (26\,$\times$\,14)\,cm$^{2}$ plastic scintillator plates separated by 10\,cm. Each plastic scintillator is read out by a photo-multiplier tube (PMT). 
Time-coincident signals in all three scintillators are interpreted as a cosmic muon event. Therefore, the number of background events from external gamma-rays or random coincidences are expected to be negligible.    
The mechanical structure allows to measure different zenith positions with an opening angle of 70\,$^{\circ}$. 

\begin{figure}
\centering
 \includegraphics[width=0.75\textwidth]{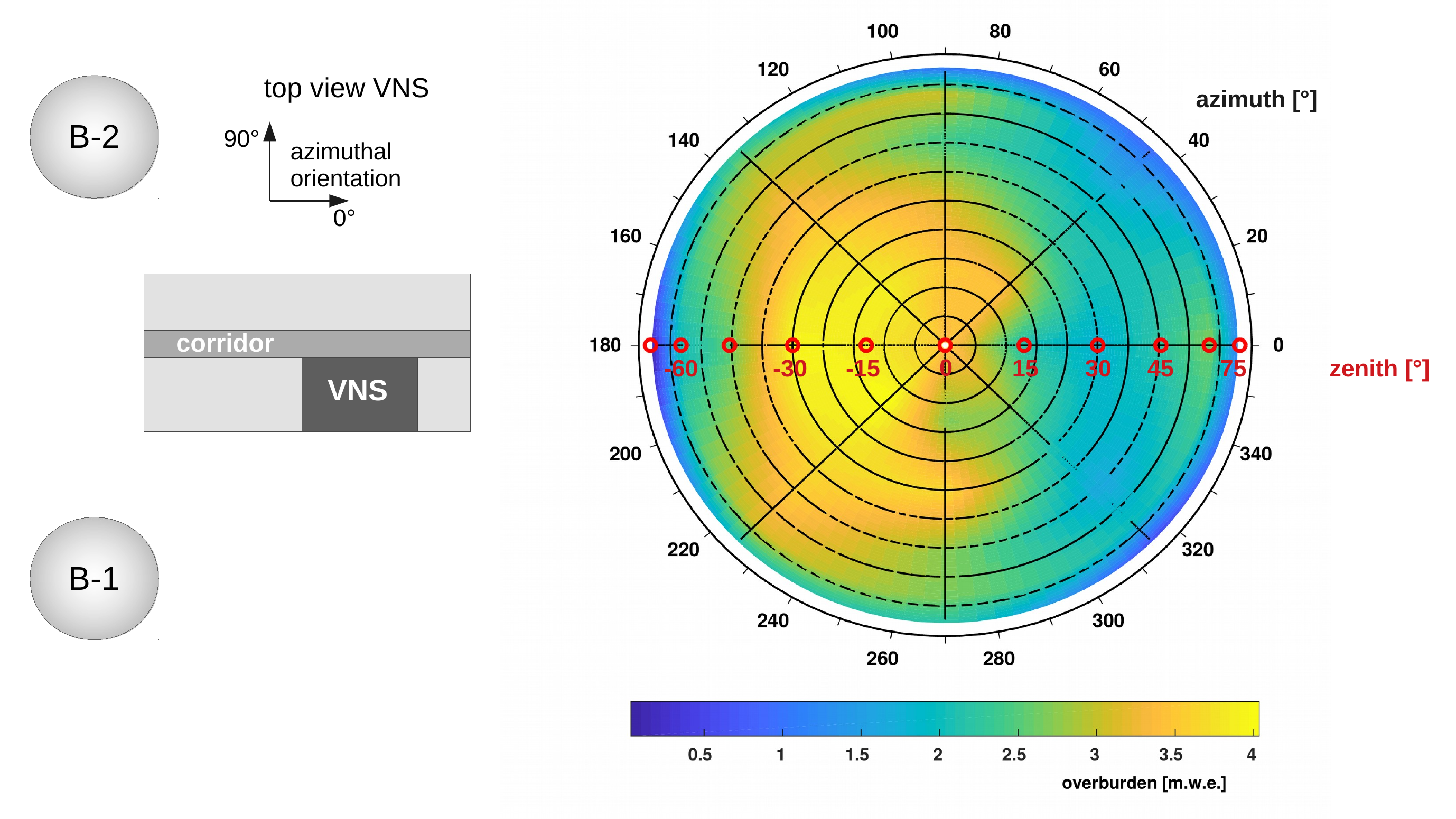}
 \caption[]{Left: Sketch of the office building hosting the VNS and the reactor buildings B-1 and B-2 as seen from the top. 
 Right: Overburden map over the VNS. The azimuth angle is measured counter-clockwise with respect to the direction opposite to the reactor buildings. The projections of the zenith angle are indicated by circles. The red markers represent the zenith angles in the measurements with $0^{\circ}/180^{\circ}$ azimuthal orientation. }
 \label{fig:overburdenmap}
\end{figure}

Figure \ref{fig:muonMeasurement} shows the measured muon rate above ground (open red markers) and four different azimuthal orientations at the VNS (filled markers).  
The angular zenithal distribution of the muon rate above ground follows the expected cos$^2\theta$ law for cosmic muons at the surface \cite{Patrignani:2016xqp}, shown by the red line. 
Based on the attenuation $a_{\mu}$ of the muon rate at the VNS, R$_{\mathrm{VNS}}$, with respect to the surface, R$_{\mathrm{surface}}$, the overburden $m_0$ is approximated with the following relation: 
\begin{equation}
 a_{\mu}\,=\,\frac{\mathrm{R}_{\mathrm{VNS}}}{\mathrm{R}_{\mathrm{surface}}}\,=\,10^{-1.32\,\mathrm{log}d\,-\,0.26\,(\mathrm{log}d)^2}
\end{equation}
where $d\,=\,1\,+\,m_0/10$, and $m_0$ is given in m.w.e. \cite{theodorsson}.

Figure \ref{fig:overburdenmap} shows a projection of the overburden over the VNS as obtained from the muon-attenuation measurements. 
Since the VNS is not centered with respect to the building, the angular distribution of the overburden is asymmetric. 
For zenith angles smaller than 50$^{\circ}$, the overburden ranges from 2.1\,m.w.e. to \,4.1\,m.w.e. 

\begin{table}[t]
\caption{Muon attenuation factor, a$_{\mu}$, at the VNS for different azimuthal angles $\phi$. $\phi$ and $\phi\,+\,180^{\circ}$ are considered as one orientation. Only statistical uncertainties are considered. }
 \centering
 \begin{tabular}{l|c|c|c|c}
  $\phi$ [$^{\circ}$] & $0/180$ & $45/225$ & $90/270$ & $135/315$\\
  \hline
  a$_{\mu}$ & 0.72\,$\pm$\,0.02 & 0.71\,$\pm$\,0.02 & 0.70\,$\pm$\,0.02 & 0.70\,$\pm$\,0.02\\
 \end{tabular}
\label{tab:MuonAttenuation}
\end{table}

For each measurement, i.e. same azimuthal orientation at the VNS and at the surface, the measured muon rates are integrated over the zenith angle to obtain $\widehat{\mathrm{R}}_{\mathrm{surface}}$ and $\widehat{\mathrm{R}}_{\mathrm{VNS,i}}$, where i\,= \,$0^{\circ}/180^{\circ}$, $45^{\circ}/225^{\circ}$, $90^{\circ}/270^{\circ}$, and $135^{\circ}/315^{\circ}$ (see Table \ref{tab:MuonAttenuation}). 
The averaged attenuation factor is given by $\overline{a}_{\mu}\,=\, 1/4\,\cdot\,\sum_{\mathrm{i}} \widehat{\mathrm{R}}_{\mathrm{VNS,i}}/\widehat{\mathrm{R}}_{\mathrm{surface}} \,=\,0.71\,\pm\,0.01$, where the error is given by the standard deviation. This  attenuation corresponds to an overburden of $m_{0}$\,=\,(2.9\,$\pm$\,0.1)\,m.w.e.

\subsection{Measurement of the Fast Neutron Flux Reduction}
\label{sec:NeutronFlux}
\begin{figure}[htbp]
\centering
\hspace{2cm}
\subfigure{\begin{overpic}[width=0.22\textwidth]{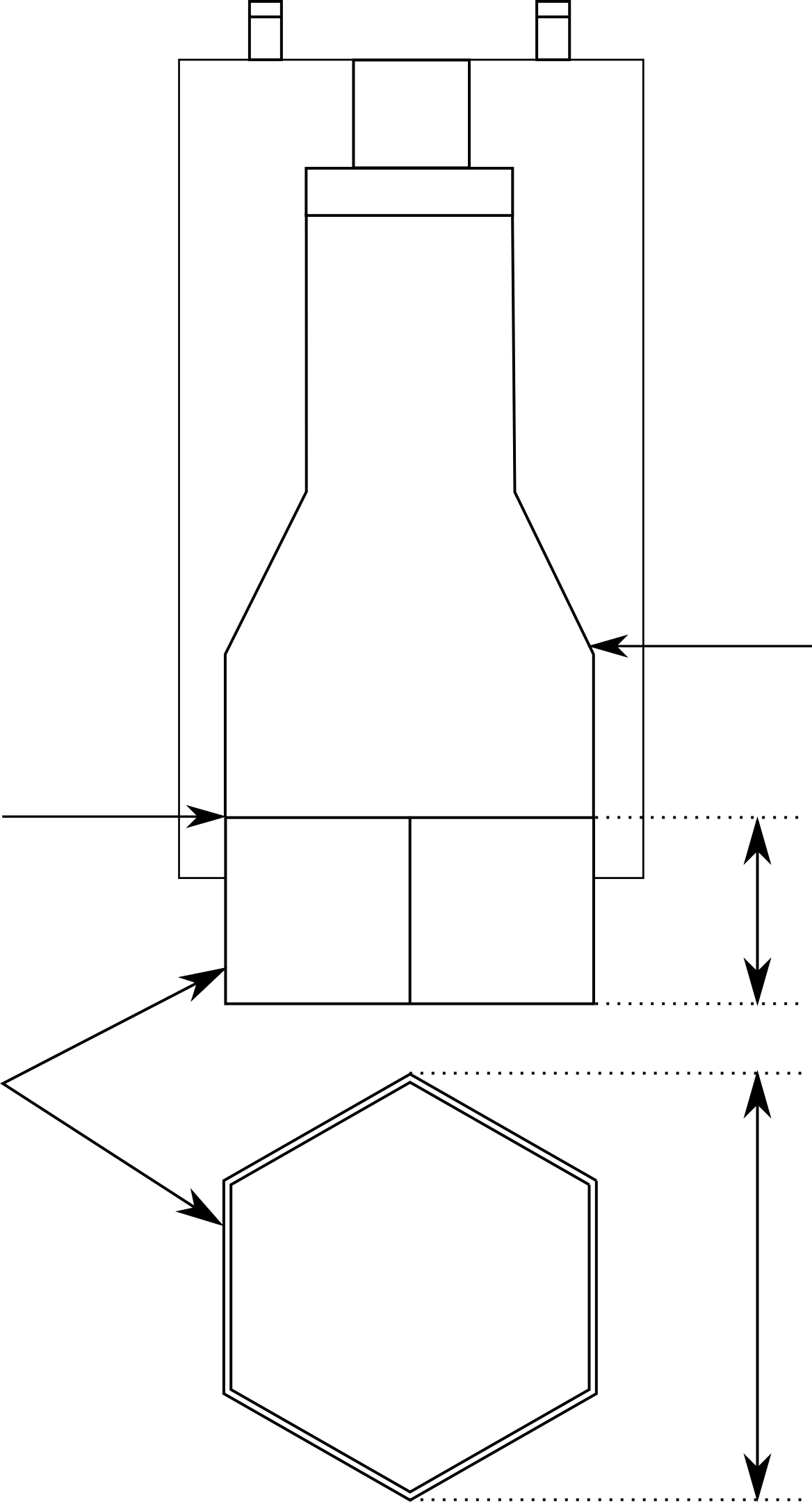}
\put (55,58){Photomultiplier}
\put (55,52){XP 3461 B}
\put (-23,45){Perspex}
\put (-32,28){Hexagonal}
\put (-32, 22){cell}
\put (55,40){\makebox(0,0){\rotatebox{90}{$\unit[50]{mm}$}}}
\put (55,15){\makebox(0,0){\rotatebox{90}{$\unit[91]{mm}$}}}
\end{overpic}}
\hspace{3cm}
\subfigure{\includegraphics[width = 0.4\textwidth]{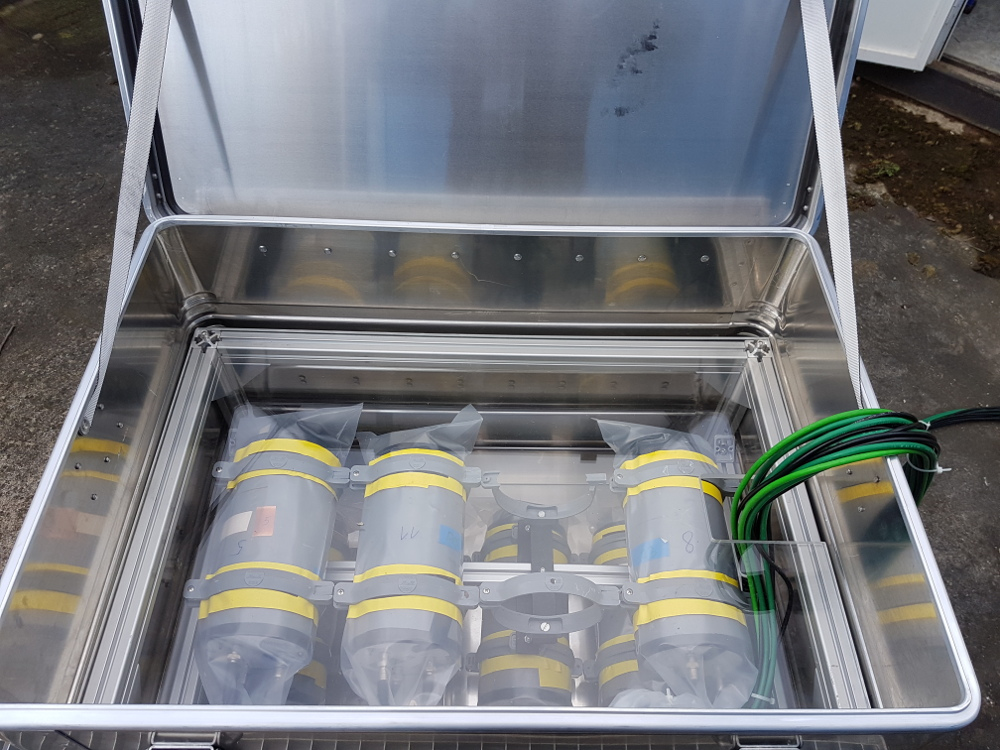}}
\caption{Left: Schematic drawing of the neutron detectors used in the neutron setup (see text). Right: Detector array box.}
\label{DetektorSkizze}
\end{figure}
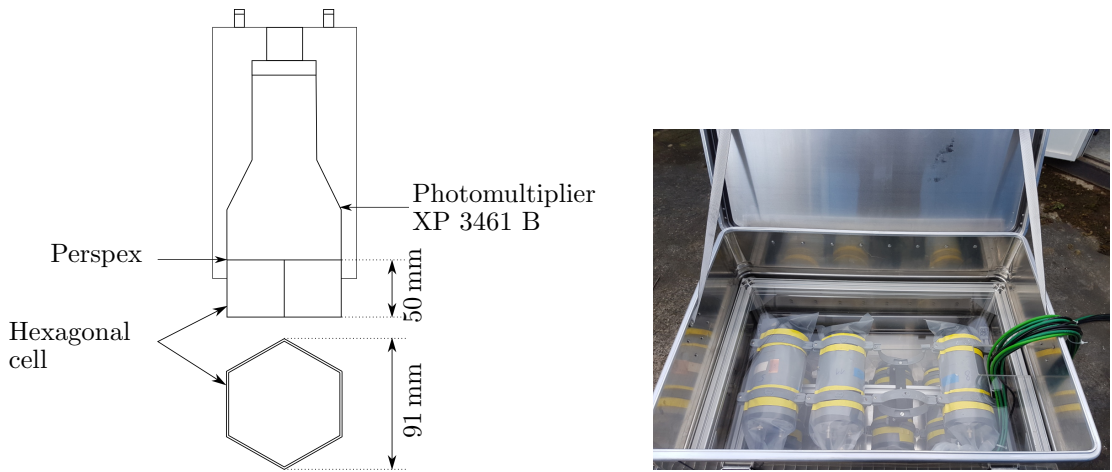

Neutrons are a potentially harmful background (discussed in Section \ref{sec:background}).
To evaluate the reduction of the neutron flux at the VNS, an array of neutron counters from Technische Universit\"at M\"unchen was deployed at Chooz. 
Each detector consists of a hexagonal cell with a diameter of $\unit[91]{mm}$ and a height of $\unit[\sim50]{mm}$. It is filled with approximately $\unit[300]{g}$ of liquid scintillator (EJ-301 from Eljen Technology \cite{Eljen}) which has a high discrimination capability between gamma-ray induced electronic and neutron-induced proton recoil. 
In order to collect the scintillation light the detectors are equipped with Philips XP- 3461-B PMTs which are optically coupled to a perspex window. Cell and PMT are enclosed in a plastic tube. A schematic drawing of the detectors can be seen in Figure \ref{DetektorSkizze}.

The setup is not only sensitive to neutrons but also to atmospheric protons. However, the flux of protons is reduced by a factor of $\sim38$ compared to the neutron flux whereby the protons are neglected in this measurement \cite{Gastrich:2015owx}. 

An array of seven such neutron detectors was operated at the Chooz power plant, at the surface and at the VNS location. The surface measurements were performed in a hut close to the Double Chooz near site (see Figure \ref{fig:Chooz}) and at negligible overburden. During the measurement campaign, the detectors were operated for a cumulative time of $\unit[15]{h}$ at the surface and $\unit[21]{h}$ at the VNS location. 

A reduced data set of three detectors, selected for discrimination threshold and performance stability, was used for the final analysis. The detectors were calibrated by identifying the Compton edges of various gamma sources ($^{22}$Na, $^{137}$Cs), wherefore the energy is given in units of electron equivalent, keV$_{ee}$, in the following.
The discrimination thresholds of the used detectors are \unit[100]{keV$_{ee}$}, \unit[400]{keV$_{ee}$} and \unit[650]{keV$_{ee}$}, respectively. 
Neutron-induced nuclear recoils in the liquid scintillator are unambiguously identified by a generic pulse-shape parameter based on the difference in the pulse decay time with respect to electron recoils. The cut is defined using an AmBe neutron calibration data set and then applied to the background data.

\begin{figure}[htb]
\centering
\includegraphics[width=0.75\textwidth]{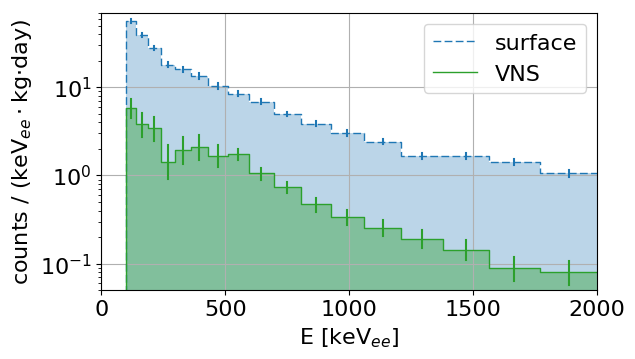}
\caption{Neutron-induced recoil rate in the liquid scintillator detectors up to an energy of \unit[2]{MeV$_\mathrm{ee}$} observed at the VNS (green histogram, solid line) and a nearby surface location (blue histogram, dashed line). For the calculation the best three detectors in terms of discrimination threshold and stability are used. Each bin displays the exposure-weighted average of the rates from all detectors with a sufficiently low threshold, along with error bars following Poissonian statistics. The histogram uses logarithmic binning with small bins at low energy to display the shape of the spectrum and large bins at high energy to account for the low statistics.}
\label{Attenuation}
\end{figure}

The data show a statistically compatible shape of the neutron recoil spectra at the VNS and surface. Thus, an energy-independent neutron reduction factor for the range \unit[100]{keV$_{ee}$}- \unit[2]{MeV$_{ee}$} is found: at the VNS the neutron flux is reduced by a factor of  $8.1\pm 0.4$ compared to the surface (see Figure \ref{Attenuation}). A detailed study of the data is in preparation.


\section{The \nucleus{} Experiment at the VNS}
\label{sec:NuCleus}
A high signal-to-background ratio is one of the key requirements for the success of a \cenns{} experiment. 
The primary \cenns{} target material in \nucleus{} will be \cawo{}. The enhanced cross-section for heavy nuclei such as tungsten, together with an envisioned detector threshold of $\mathcal{O}(10\,\mathrm{eV})$, will give a strong \cenns{} signal (see Figure \ref{fig:RatePlot} and Equation \ref{eq:cenns_xsection}). 
One of the main experimental challenges is to achieve a sufficiently low background level in the \cenns{} target detector. To reach the desired background count rate of the order of $10^2$\,\rateunit{}, the \nucleus{} experiment will apply passive and active background suppression methods. 

In the following section, the potential background sources for the \nucleus{} experiment at the VNS are discussed. 
A first concept of the \nucleus{} experiment is presented in Section \ref{sec:NucleusDetectors} with focus on active and passive background suppression. 
The BASKET R\&D project, presented in Section \ref{sec:basket}, gives the opportunity to complement the \cenns{} detection concept of \nucleus{} with a unique idea to characterize the neutron background, one of the most challenging backgrounds for \cenns{} experiments.

\subsection{Sources of Background}
\label{sec:background}
Experimental sites close to nuclear reactors are usually located at shallow depth, typically corresponding to an overburden of $\mathcal{O}(10\mathrm{\,m.\,w.\,e.})$. 
The background sources of interest for the \nucleus{} experiment are here  categorized into external, internal, cosmic-ray induced and reactor-correlated backgrounds. 
The dominant sources are briefly summarized below, for a more detailed description the reader is referred to \cite{Heusser:1995wd,Formaggio2004}.

The external background is dominated by gamma-rays from $\alpha$- and $\beta$-decaying nuclides in the $^{232}$Th, $^{238}$U and $^{235}$U decay chains, as well as $^{40}$K which is present in material surrounding the experiment. 
Gamma-rays originating in the concrete of the laboratory building can be shielded by high-Z material, such as lead.   
Materials used for components close to the target detectors, such as mounting and support structures or electronic components, need to be carefully selected, e.g. by exploiting $\gamma$-ray spectroscopy, and undergo cleaning and purification processes \cite{Formaggio2004}. 

Nuclides from the thorium and uranium decay chains may also be present in the detector target itself. In the case of \nucleus{}, this internal contamination can be reduced e.g. by special measures during crystal production \cite{Erb2013}.
In the uranium and thorium decay chains, radon is produced which escapes by diffusion or nuclear recoil accompanying the ejection of an $\alpha$-particle. 
Radon is the strongest source of airborne radioactivity \cite{Formaggio2004}.
To prevent radon deposition on the detector surface, special cleaning procedures are needed during detector assembly and installation.

The above-mentioned decays typically produce $\gamma$-rays with energies up to 2.6 MeV, which is the highest naturally occurring $\gamma$-line from the decay of $^{208}$Tl (a progeny of $^{232}$Th). $\gamma$-rays of MeV energies predominantly interact via Compton scattering, thus the energy deposited in the target detector ranges from zero up to the energy of the Compton edge. 
Since $\gamma$-rays likely interact via multiple Compton scattering, a fiducialization of the \nucleus{} detector is a promising tool to actively suppress the external $\gamma$-background (see Section \ref{sec:NucleusDetectors} for further details).

The energies of $\alpha$- and $\beta$-particles typically reach up to several MeV. 
In $\beta$-decays, the available energy is split between the emitted $\nu$ and the $e^{+/-}$.  
While a $\beta$-particle is typically absorbed within a few mm in a solid, the neutrino remains undetected. If the $\beta$-decay occurs on the surface of the detector or is followed by the emission of $\gamma$-rays, the fiducialization of the detector volume allows to discriminate this background from \cenns{} events. 
The range of an $\alpha$-particle does not exceed a few tens of $\mu$m in a solid. 
If released in the bulk of the target detector, $\alpha$-particles are fully absorbed, and the event exhibits an energy well above the (sub-)keV signal of a \cenns{} event. 
However, if the $\alpha$-decay happens on or close by the detector surface a significant fraction of the energy remains undetected and such an energy deposit may mimic a (sub-)keV event.
By instrumenting the surrounding material surface events can be discriminated from \cenns{} events (see Section \ref{sec:NucleusDetectors}). 

Most of the secondary cosmic rays from air showers like electrons, $\gamma$-rays, protons and pions are absorbed by the building structure or a dedicated lead shielding.
The remaining muons and neutrons are the main components of cosmic-ray-induced background at shallow overburden \cite{Heusser:1995wd}.
Besides a source of background via ionization, pair production and Bremsstrahlung, muons also produce neutrons in nuclear reactions, especially in high-Z materials like lead.
The flux of neutrons caused by air showers decreases rapidly with increasing overburden. 
Already with an overburden of 3\,m.w.e. the neutron flux starts being dominated by neutrons produced by muons in a massive lead shielding \cite{Heusser:1995wd}.
The experimental signature of a neutron is a nuclear recoil, and thus, neutrons are a particularly dangerous source of background for any \cenns{} experiments.
Furthermore, they can be highly penetrating and difficult to shield. Neutrons may also produce radio-nuclides by inelastic scattering and neutron-capture in the detector or surrounding materials. 

To reduce the muon-induced background, an active muon veto, typically based on plastic scintillators is used: all events in coincidence with a muon event are discarded. 
In this way, muon-induced events, which are e.g. induced by ionization or Bremsstrahlung, or by neutrons produced in the shielding material, are vetoed.  
Neutron-induced radio-isotopes with lifetimes much shorter than the coincidence window are rejected as well. 
In order to decrease the time for thermalization and capture of neutrons in the shielding, moderators, e.g. polyethylene (PE), are used. 
To attenuate neutrons produced inside the shielding, it is most efficient to place a neutron absorber inside the high-Z shielding material \cite{Heusser:1995wd}. 

Nuclear reactors emit a large number of $\gamma$-rays and neutrons. $\gamma$-rays are produced in the fission process as well as the subsequent $\beta$-decay, neutron capture or de-excitation  of the fission products. High energy $\gamma$-rays may produce neutrons in photo-nuclear reactions e.g. in the concrete of the reactor building \cite{Bui:2016otf}. 
The VNS is located at a distance of more than 70\,m from the cores, with $\mathcal{O}$(10\,m) of rock and concrete in the line of sight. Furthermore, the reactor vessel is shielded by a thick layer of steel. Therefore, $\gamma$-ray induced neutron background is negligible. 

Fission induced neutrons are emitted with a mean energy of 2\,MeV. 
Most of these primary neutrons are thermalized in the reactor. 
The expected reactor-correlated fast neutron background at the VNS is estimated referring to the experience obtained with the NUCIFER \cite{Boireau:2015dda}
experiment, a project designed for online monitoring of nuclear reactors. 
The NUCIFER detector was based on Gd-loaded liquid scintillator and was operated 7.2\,m from the core of the Osiris research reactor (70\,MW) in Saclay, France.  
The estimated neutron elastic scattering rate on hydrogen in the whole NUCIFER target volume was $4\,\cdot\,10^{-5}\,\mathrm{events}/(\mathrm{day})$  for energies above 2\,MeV \cite{Boireau:2015dda}.
Scaling this result to the relevant parameters of \nucleus{}, i.e. reactor power of 2$\cdot$4.25\,GW$\mathrm{th}$, distance of 72\,m between the detector and the core, and assuming the same scattering cross-section on each nuclide of CaWO$_4$ as on hydrogen, indicates that the neutron background from the reactor core is negligible.

\subsection{The \nucleus{} Experiment}
\label{sec:NucleusDetectors}
\begin{figure}
\centering
\includegraphics[width=0.8\textwidth]{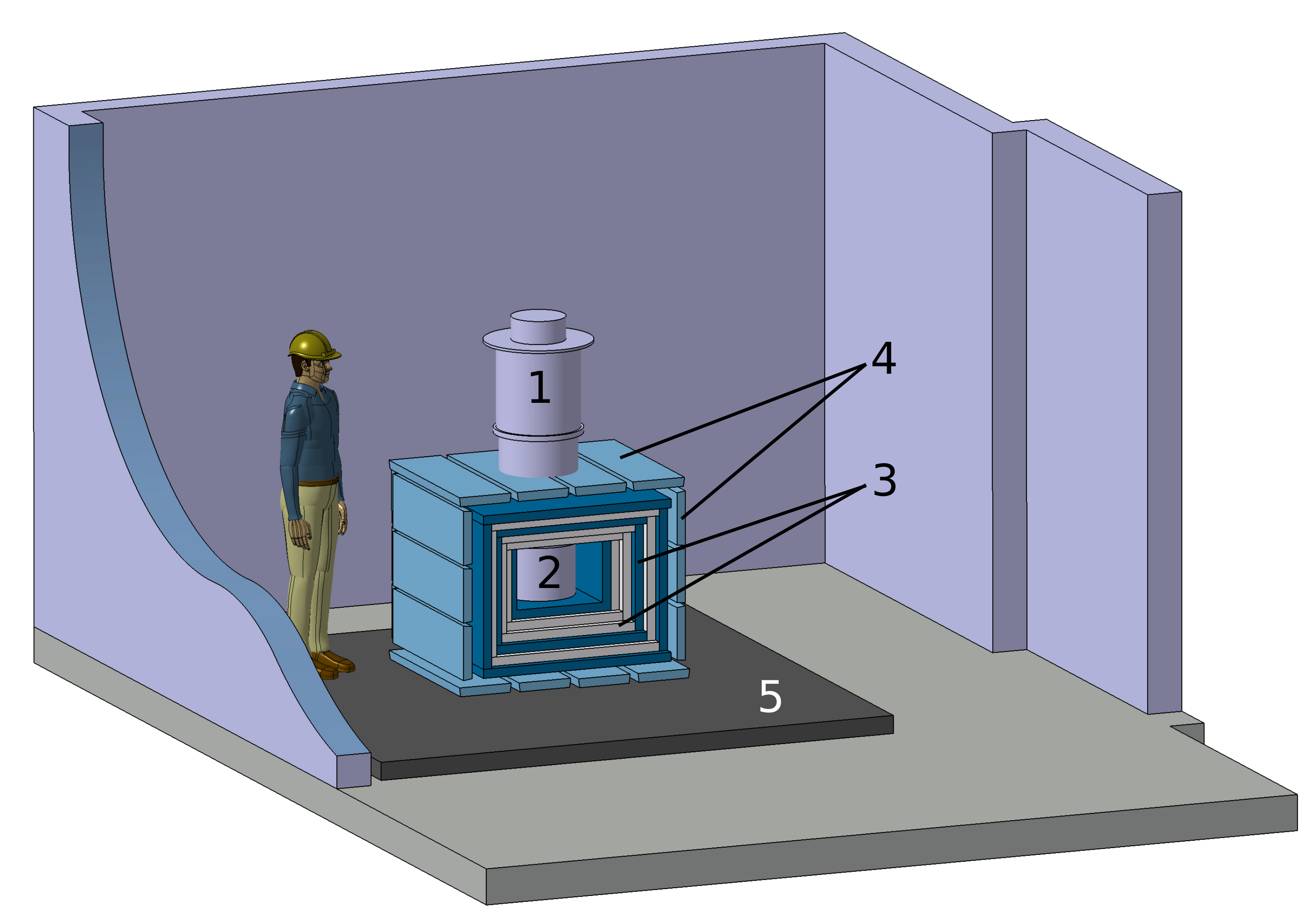}
\caption{Sketch of the \nucleus{} experiment at the VNS. The setup consists of a cryostat (1), where the target detectors are installed in its lower part, the experimental volume (2). The latter is surrounded by a passive shielding (3) consisting of alternating layers of borated PE and lead. The top part will be completed by a cold shield inside the cryostat. The outermost layer is the active muon-veto (4) made from plastic scintillator panels. 
The full setup with a footprint of approximately 1\,m$^2$ is placed on a weight support platform (5). Further support structures, e.g. for the cryostat, are omitted for clarity. }
\label{fig:Setup}
\end{figure}

A promising technology for the study of \cenns{} are cryogenic calorimeters.
Cryogenic detectors measure the temperature rise $\Delta\mathrm{T}\,=\,\Delta\mathrm{E}/\mathrm{C}$ following an energy deposition $\Delta\mathrm{E}$ in a target with a heat capacity C. 
For crystalline material kept at a temperature $\mathcal{O}(\mathrm{mK})$, C is small enough to achieve a measurable $\Delta\mathrm{T}$ even for small energy depositions.   
The detector technology of \nucleus{} is based on CRESST cryogenic detectors \cite{Angloher2016}
which have world-leading sensitivity in the field of direct low mass dark matter search and are driven by similar requirements as the study of \cenns{}: a sub-keV energy threshold as well as a low intrinsic background. 
The CRESST experiment has developed \cawo{} and \alo{} crystals as targets with masses between 24\,g and 300\,g. 

The \nucleus{} concept expands this technology to an ultra-low-threshold ($\mathcal{O}$(10\,eV))  gram-scale cryogenic calorimeter~\cite{PhysRevD.96.022009} as a target detector for \cenns{} combined with cryogenic veto detectors for active background discrimination. The experiment will proceed in two phases: in a first step, \nucleus{}-10g, a detector with a 10\,g \cenns{} target will be deployed. A later stage, \nucleus{}-1kg, foresees an upgrade to a total target mass of 1\,kg. 
Figure \ref{fig:Setup} shows a first design of the proposed \cenns{} experiment at the VNS. The \nucleus{} target detectors are installed in a cryostat. The experimental volume is surrounded by a shielding for active and passive background reduction.

\subsubsection{The \nucleus{} Target Detectors}
\begin{figure}[htbp]
\centering
\includegraphics[width=0.8\textwidth]{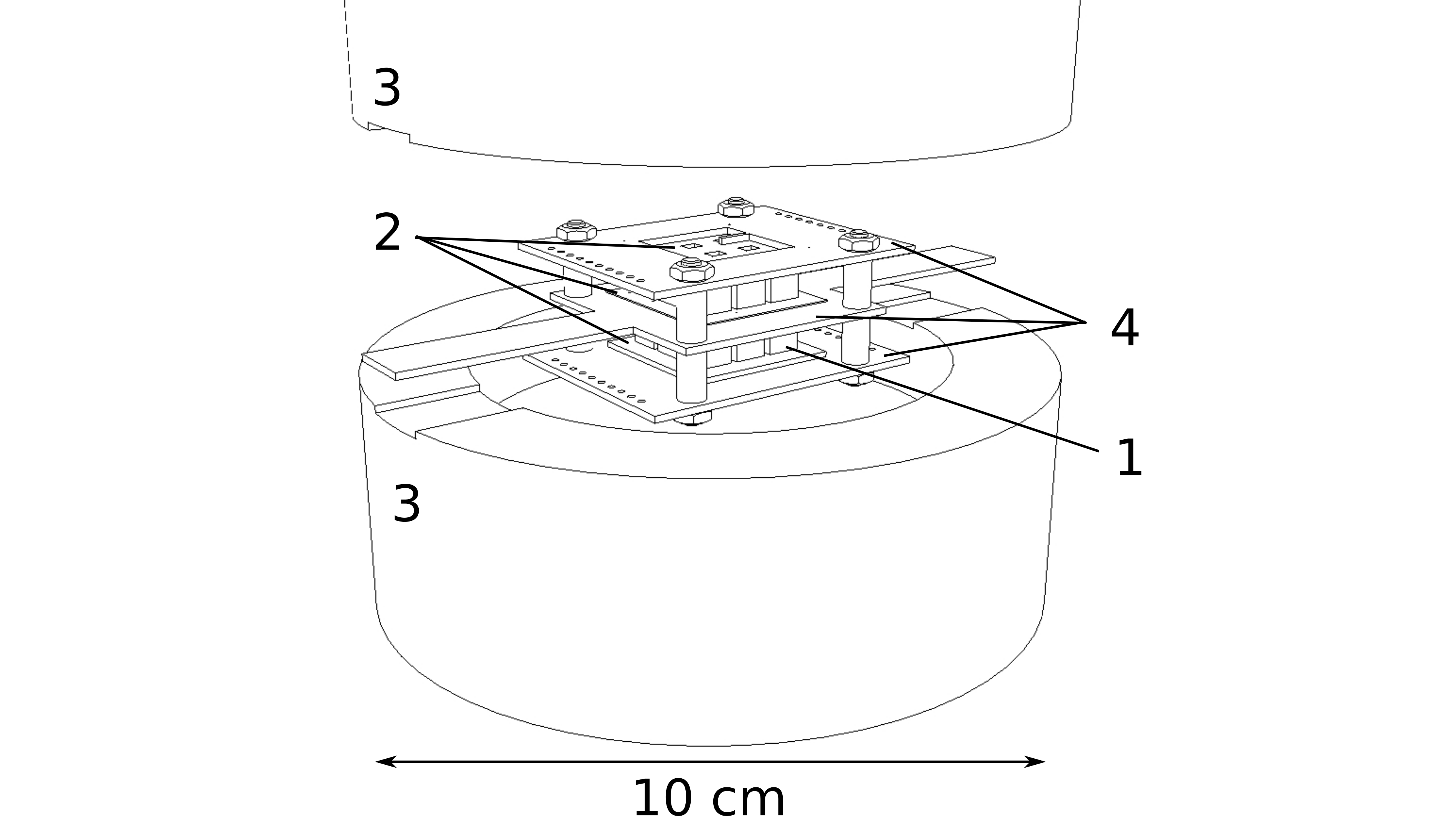}
\caption{3D sketch of the \nucleus{}-10g detector. It consists of three different types of cryogenic calorimeters -- two 3$\times$3 arrays of gram-scale cryogenic calorimeters as \cenns{} target (1), an inner veto (2) and an outer veto (3) with a diameter of 10\,cm. The assembly is held mechanically by a non-instrumented support structure (4). The target is operated in anti-coincidence with the inner and the outer veto. See text for details. 
}
\label{fig:NucleusSchematic}
\end{figure}

The \nucleus{} detector concept is a fiducial-volume cryogenic detector (see Figure \ref{fig:NucleusSchematic}) based on three individual calorimeter systems: (1) an array of gram-scale cryogenic calorimeters with ultra-low energy threshold as target detector, (2) a low-threshold inner veto holding and encapsulating the target, and (3) a surrounding kg-scale cryogenic detector used as outer veto. 

CaWO$_4$, Al$_2$O$_3$, Si and Ge are well known materials for cryogenic detectors and suitable candidates for the target volume of the experiment. 
Detectors with a total mass of $\mathcal{O}(\unit[1]{g})$ enable ultra-low thresholds \cite{PhysRevD.96.022009}
and imply a total event rate per crystal of $\lesssim$\unit[1]{Hz} which is crucial to keep the deadtime at a value of $\mathcal{O}(1\%)$. 
In the CRESST experiment \cite{Angloher2016}, transition edge sensors (TESs) have proven to be a highly sensitive thermometer able to measure the low recoil energies produced in \cenns{}.
With a 0.5\,g prototype detector made from a (5\,mm)$^3$ Al$_2$O$_3$ cubic crystal, an unprecedented ultra-low threshold of E$_{\mathrm{th}}$\,=\,(19.7\,$\pm$\,0.9)\,eV has been reached \cite{Strauss:2017cuu}, one order of magnitude lower than previous devices. Such gram-scale crystals exploit the full potential of the \cenns{} signal (see Figure 1). Arrays of these gram-scale cryogenic calorimeters allow deploying a larger total target mass. Figure \ref{fig:NucleusSchematic} shows a schematic view of \nucleus{}-10g: a 3$\times$3 array of CaWO$_4$ (6\,g) and a 3$\times$3 array of Al$_2$O$_3$ (4\,g). This multi-target approach will provide an in-situ background characterization: 
while the \cenns{} scattering rate ($\sim$N$^2$) is strongly enhanced for CaWO$_4$, fast neutrons are expected to induce comparable signatures in CaWO$_4$ and Al$_2$O$_3$ due to scattering on O nuclei \cite{Strauss:2017cuu}. Furthermore, coincident background events in several target detectors can efficiently be discriminated from \cenns{} events, which are single scatters.

The target detectors will be embedded within cryogenic veto detectors which allow for an efficient active background reduction. The \nucleus{} detector support structures are made of Si-wafers read out by TESs which enable a 4$\pi$ low-threshold inner veto. 
Background events originating from surface $\alpha$ and $\beta$ contamination of the detector components can be rejected by anti-coincidence~\cite{Strauss:2017cuu}.
Both target and inner veto will be enclosed in a kg-scale outer veto which is operated in anti-coincidence mode to reject neutron and $\gamma$-ray backgrounds. Ge and CaWO$_4$ are established materials for cryogenic detectors with masses up to $\unit[1]{kg}$ and typically reach energy thresholds of a few keV. 
Ge single crystals are commercially available in extremely high purity and large sizes.
CaWO$_4$ is especially interesting due to the presence of tungsten with a high cross section for $\gamma$-radiation, and oxygen which enables efficient neutron moderation. An outer veto from \cawo{} would require R\&D into large-diameter crystal production.
Studies based on Monte Carlo simulations show that with the outer and inner vetos a background suppression of $\mathcal{O}(10^3)$ can be achieved \cite{Strauss:2017cuu}. 

The \nucleus{} detector concept foresees to scale the total target mass from gram to kilogram. 
The production of the cryogenic detectors is based on well-established techniques of the semiconductor industry. In particular, the capability to produce multiple target elements simultaneously before cutting and polishing has been demonstrated,
which is crucial for the later stages of \nucleus{}. 

\subsubsection{The Cryostat}
The base temperature of $\mathcal{O}$(10\,mK) required to operate cryogenic detectors is typically provided by ``wet'' dilution refrigerators relying on a continuously evaporating bath of liquid helium at 4\,K. At the VNS, the handling of cryogenic liquids can be avoided using a ``dry'' dilution refrigerator which replaces the helium bath with a pulse-tube cryocooler integrated into the cryostat. Though much simpler in operation and infrastructure requirements, this feature makes a ``dry cryostat'' a challenging environment for cryogenic detectors due to vibrations induced by the pulse-tube, which will degrade the detector performance if unaddressed. This challenge can be met by dedicated work on vibration decoupling~\cite{Maisonobe:2018tbq}, which is mandatory for the \nucleus{} setup.
The \cenns{} target detectors are installed in the experimental volume inside the cryostat. The cryogenic infrastructure is chosen such as to host both \nucleus{} phases. The experimental cryogenic volume is  surrounded by a compact passive and active shielding.

\subsubsection{Active and Passive Shielding}
\label{sec:APShielding}
Many examples in literature demonstrate that a low background count rate can be reached with a shallow overburden of $\mathcal{O}(10\,\mathrm{m.w.e.})$.
Two promising examples are GIOVE, a highly sensitive HPGe spectrometer in the shallow depth laboratory (15\,m.w.e.) at the Max-Planck-Institut f\"ur Kernphysik \cite{Heusser:2015ifa}, and the Dortmund low-background facility with an overburden of 10\,m.w.e \cite{Gastrich:2015owx}. Both set-ups feature an active muon-veto and passive shields of lead, copper, and borated PE. With these graded shielding structures background rates of 0.13\,counts/(keV$\cdot$kg$\cdot$d) for GIOVE and 1.4\,counts/(keV$\cdot$kg$\cdot$d) for the Dortmund low-background facility are reached in the 40-2700\,keV energy range.

The goal of the \nucleus{} experiment is to achieve a background count rate of $10^{2}\,\mathrm{counts}/(\mathrm{keV}\cdot\mathrm{kg}\cdot\mathrm{d})$ in the sub-keV region. At present, this energy range has never been probed in a shallow depth laboratory.
Based on the experience of the aforementioned experiments operating in the keV-region, the passive  shielding of \nucleus{} will consist of several alternating layers of borated PE and lead. To avoid a hole directly above the detectors, a part of the shielding has to be mounted inside the cryostat and kept cold during operation.
Neutrons produced in nuclear reactions are moderated by the PE and captured on boron. 
External gamma-rays and such produced in radiative neutron captures (for boron the most intense line is at 478\,keV \cite{iaea2007})
are attenuated by the lead. 
As copper can be produced with very low intrinsic radioactive contamination 
compared to lead (see e.g.\ \cite{Alvarez2013}), copper may be used as innermost shielding layer. 
The cubic passive shielding will have an approximated edge length of 1\,m. 
Its final dimension and layout, i.e. the sequence and thickness of the PE and Pb layers, as well as the boron doping of PE, will be based on dedicated Monte Carlo studies and is the subject of future publications. 

As discussed in Section \ref{sec:background}, cosmic-ray induced events are an unavoidable source of background. Cosmic muons cannot be sufficiently attenuated by a passive shielding, thus, muon-induced background events need to be identified and removed from the data. 
An active muon veto consisting of 5\,cm thick plastic scintillator plates is planned. 
With an energy loss of $\sim2\,\mathrm{MeV}/\mathrm{cm}$ a 4\,GeV muon (mean energy of the cosmic muons at the surface) deposits about  10\,MeV in a veto panel \cite{Patrignani:2016xqp}.
The most energetic gamma-ray from natural radioactivity is at 2.6\,MeV, see Section \ref{sec:background}. 
Thus, cosmic muons can efficiently be discriminated by a simple energy cut. 
An event in the muon-veto that exceeds the energy threshold, causes a muon-trigger. 
Events in the target detector in coincidence with a muon trigger are disregarded.

\subsection{The BASKET Detectors}
\label{sec:basket}
BASKET (Bolometers At Sub-KeV Energy Threshold) is an R\&D program started in 2017 at CEA in collaboration with CSNSM  (Centre de Sciences Nucléaires et Sciences de la Matière) to develop innovative detectors for neutrinoless double beta decay and \cenns{} \cite{basket:irfupage}.  
For the latter, the project focuses on the development of Li$_2$WO$_4$ crystals as a new absorber material together with new thermal sensors to optimize the time response and energy resolution.
First tests on an 11-g crystal, which was read out with a Neutron Transmutation Doped Ge sensor and a Ge Neganov-Luke light detector, showed that Li$_2$WO$_4$ exhibits good bolometric and scintillation properties.
Furthermore, it could be demonstrated that $\alpha$-particles can be discriminated from $\beta$-particles and $\gamma$-rays with a discrimination power of more than five sigma \cite{PhDZolotarova}. 
Further tests with 1-g crystals coupled to metallic magnetic calorimeters \cite{Gray2016} and NbSi TES \cite{Crauste2011} are currently performed. 
The goal is to reach an energy threshold of $\mathcal{O}$(10\,eV) and a time response of $\mathcal{O}$(100\,$\mu$s). 
A possible synergy of BASKET with the \nucleus{} experiment is the deployment of gram-size Li$_2$WO$_4$ crystals as target detectors. The \cenns{} rate is similar to the one expected for CaWO$_4$. Moreover, an additional target compound to CaWO$_4$ and Al$_2$O$_3$ may yield supplementary information on the background yet to be explored at a few tens of eV. 

Furthermore, Li$_2$WO$_4$ is a very interesting material to monitor neutron backgrounds and could be used for the outer veto of \nucleus{}. The heavy element tungsten has a high attenuation power for $\gamma$-rays. Neutrons can be tagged using the neutron capture on $^6$Li which has a natural abundance of about 7.4\%, allowing for an in-situ characterization of the neutron background \cite{lmo:bekker2016}. 
The reaction $^6$Li + n $\rightarrow$ $^3$H + $^4$He, with an energy release of E\,=\,4.78\,MeV\,+\,E$_{\mathrm{n}}$, where E$_{\mathrm{n}}$ is the kinetic energy of the neutron, produces only heavy charged particles. The latter are well separated from electronic recoils ($\beta$, $\gamma$, $\mu$) by the reduced light signal. 
Thus, the identified $^{6}$Li neutron capture events can be used to constrain the neutron rate and allow fast neutron spectroscopy.
In order to increase the capture rate, isotopically enriched $^6$Li$_2$WO$_4$ may be used. 
Currently, techniques to grow large-sized Li$_2$WO$_4$ crystals as well as the possibility of enrichment are under investigation.


\section{Muon Induced Dead Time Considerations}
\label{sec:DeadTime}
To reach the required background level for \nucleus{} at a shallow experimental site, a compact passive shielding in combination with an efficient muon veto is compulsory (see Section \ref{sec:APShielding}). 
At the Earth's surface the count rate of cosmic muons is $\sim$100\,Hz/m$^2$ \cite{Patrignani:2016xqp}. 
Hence, to avoid significant detector dead time the timing of the coincident events has to be sufficiently fast. 
While typical muon-veto panels feature pulse rise times $<1\,\mu$s,  pulses of cryogenic detectors as the one used for \nucleus{} are typically orders of magnitude slower which makes the detector operation at shallow depth challenging. 

\subsection{Monte Carlo Simulations of Cosmic Muons at the VNS}
\label{sec:MCcosmics}
A Monte Carlo (MC) simulation tool based on GEANT4 \cite{Agostinelli:2002hh} was used to simulate cosmic muons at the VNS and to estimate an upper limit on the trigger rate of the muon-veto. 
Muons are randomly generated on a plane tangent to a half-sphere centered at the detector whereas the distribution of the point of tangency follows the cos$^2\theta$-distribution of muons on the surface. The direction of the generated muons is parallel to the normal vector of the plane, pointing towards the detector.   

To validate the MC simulations, a shielding configuration similar to the one
used in the NUCIFER experiment \cite{Boireau:2015dda} was simulated and the
amount of simulated hits in the shield was compared to the published
experimental data. 
The differential flux of incident atmospheric muons $d\Phi/dE
d\Omega$ is approximated according to an adapted Gaisser parametrization \cite{Gaisser1991, Chirkin2007}. 
For simplicity, only mono-energetic muons between 4.5\,MeV and 450\,GeV in
logarithmic binning were simulated. The results were weighted according to the
flux within the given energy range with respect to the total flux.
The trigger rate of the muon-veto is calculated by
\begin{equation}
 \mathrm{R}\,=\,\frac{\mathrm{N}_{\mu}}{\mathrm{N}_{\mathrm{MC}}}\cdot 2\pi\cdot \Phi_{\mu}\cdot\mathrm{A}\cdot \mathrm{a}_{\mu}
 \label{eq:MuonRate}
\end{equation}
where $N_\mathrm{MC}$ is the number of generated muons in the simulation,
$N_\mu$ is the number of energy depositions in the muon-veto (i.e. a muon
trigger) which pass the 10 MeV threshold \cite{Boireau:2015dda} of the NUCIFER
experiment,
$\Phi_{\mu}$\,=\,$70/(\mathrm{sr}\cdot\mathrm{s}\cdot\mathrm{m}^{2})$ is the
total muon flux at the surface \cite{Patrignani:2016xqp}, $A=(3.5\,\mathrm{m})^2$ is
the surface of the plane where the muons are generated, and $a_\mu$ is the
muon-attenuation at the experimental site.
In total $10^{4}$ muons were generated for each muon energy bin.

\begin{table}[]
    \centering
     \caption{Number of simulated muon triggers, N$_{\mu}$, for the NUCIFER experiment for different muon energies E. In total 10$^{4}$ muons were simulated for each energy. The weights are given within the logarithmic energy bin centered at the energy E with respect to the total flux. The total number of triggers is given by the weighted sum of the mono-energetic simulations. }
    \begin{tabular}{l|c|c|c|c|c|c||c}
    \toprule
        Energy [GeV] & 0.0045 & 0.045 & 0.45 & 4.5 & 45.0 & 450.0 & total \\
        \midrule
         weight [\%] & 0.28 & 2.73 & 22.03 & 58.27 & 16.22 & 0.48 & 100 \\
         N$_{\mu}$ & 0 & 2 & 2580 & 2816 & 2817 & 3027 & 2680.43 \\
         \bottomrule
    \end{tabular}
    \label{tab:NUCIFER_results}
\end{table}

Table \ref{tab:NUCIFER_results} summarizes the simulated muon trigger rates in
the NUCIFER setup.
The muon simulations with the simplified NUCIFER geometry yield a total muon
trigger rate of (535\,$\pm$\,9)\,Hz, by taking into account a veto-efficiency of
97\% the trigger rate is reduced to (519\,$\pm$\,9) Hz.
The MC result is about 50\% higher than the measured muon trigger rate of
350\,Hz \cite{Boireau:2015dda}.
This deviation may originate in the simplifications of the detector geometry,
and in further unaccounted inefficiencies.
It is believed that the applied approximations to the muon energy spectrum give the largest systematic uncertainty. The simulation assumes a uniform overburden of 12\,m.w.e, and does not account for any changes in the muon energy spectrum with respect to the surface. 
Furthermore, the rough binning may be over-simplified.

\begin{figure}
 \centering
 \includegraphics[width=0.5\textwidth]{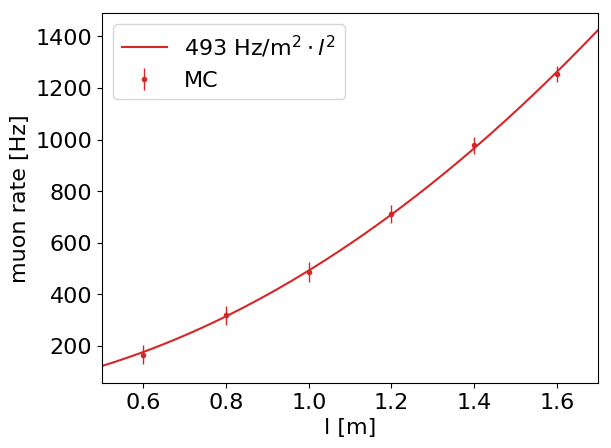}
 \caption{Simulated rate of incident muons with 4\,GeV energy for a cubic muon-veto at the VNS with varying edge length $l$. The rate shows a quadratic increase with the edge length (red line). Uncertainties are  statistical only.}
 \label{fig:MuRate_vs_VetoSize}
\end{figure}

As the goal is to estimate an upper limit on the trigger rate for the actual
\nucleus{} muon-veto, a verified overestimation of 50\% in the simulated trigger rate is not an obstacle.
To estimate the muon rate, a simplified cubic geometry with six plates, i.e. at
the top, bottom and four side faces, of equal size was implemented in the
simulation tool. Each plate of the muon-veto is made from standard PE
(C$_2$H$_4$) with a density of 0.96\,g/cm$^3$.
For simplicity the generated muons have an energy of 4\,GeV which corresponds to
the mean energy of cosmic muons at the surface. For the muon attenuation, a
uniform factor of a$_{\mu}$\,=\,0.71 is assumed, see Section \ref{sec:MuonAttenuation}.
No trigger threshold is applied, which is legitimate as only an upper 
limit on the trigger rate is estimated. 
The muon rate is calculated using Equation \ref{eq:MuonRate}.
As seen in Figure \ref{fig:MuRate_vs_VetoSize}, the rate is proportional to the
surface of the muon-veto.
For a shielding covering a volume of 1\,m$^3$, as envisioned by \nucleus{}, the simulation yields an upper limit on the  muon trigger rate of (487\,$\pm$\,38)\,Hz.

\subsection{Estimation of Muon-Induced Dead Time}
\label{sec:DTEstimation}
The overall detector dead time is governed by the timing of the cryogenic pulses, since the timing of the muon-veto panels is orders of magnitude faster and, therefore, can be neglected. 
In the following, the pulse timing, i.e. the uncertainty $\sigma_\tau$ of the determination of the pulse's onset 
is assumed to be proportional to the rise time.
In cryogenic detectors, the rise time is given by the thermalization of the signal phonons at the surface of the absorber crystal, and is thus expected to scale roughly with the linear dimension of the target. 
The pulse rise time of the 0.5\,g \nucleus{} detector prototype is $\tau_r\approx300\,\mu$s \cite{PhysRevD.96.022009}, significantly faster than CRESST-II detectors with a mass of 300\,g and a typical rise time of several ms~\cite{PhDStrauss}, and with other cryogenic detector technology.

For the estimation of the detector dead-time, a time window of $\pm5\sigma_\tau$ is defined around every event in the muon-veto and removed from the total exposure time.  
For example, in case of a cubic muon-veto of 1\,m side length with a count rate of $R=(487\,\pm\,38)$\,Hz the pulse's onset has to be determined with $\sigma_\tau\lesssim20\,\mu$s to restrict the dead time to $\lesssim10\%$.
The precision of the pulse onset of a fast cryogenic detector with a rise time of 100\,$\mu$s 
has been measured using a pulsed neutron beam from an accelerator in \cite{Strauss:2014zia}. 
The uncertainty of the onset determination is 
$\sigma_\tau=(4.8\pm0.4)\,\mu$s corresponding to a time window of $48\,\mu$s around every event in the muon-veto to be removed from the exposure time.

\begin{figure}
\centering
 \includegraphics[width=.7\textwidth]{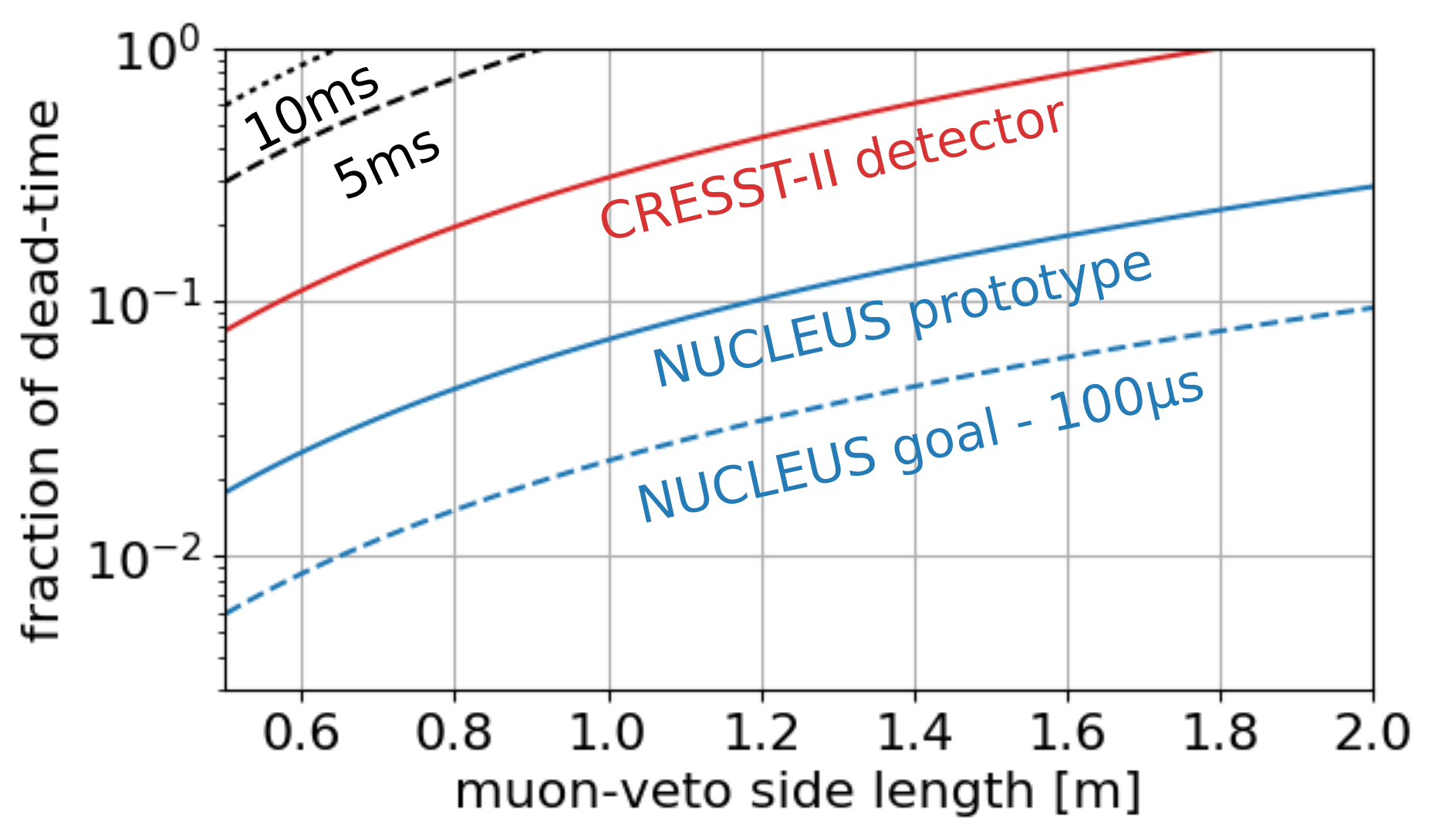} 
 \caption{ Detector dead time vs. dimension of the cubic-shaped muon-veto. The lines indicate the predicted values for the  \nucleus{} prototype detector (300\,$\mu$s rise time, blue line) and a CRESST-II detector (1.3\,ms rise time, red line) based on TES technology. The relation between rise time and timing uncertainty is taken from a neutron beam coincidence measurement (see text). For comparison, expected dead-times of slower detectors with 5\,ms rise time (black dashed line) and 10\,ms rise time (black dotted line), as well as a \nucleus{} detector with an envisioned rise time of 100\,$\mu$s (blue dashed line) are shown. }
 \label{fig:deadTime}
\end{figure}

Using the results of the dedicated MC simulation (see section \ref{sec:MCcosmics}) for the muon count rate of different shielding dimensions, the resulting dead time can be calculated with respect to the side length of the cubic-shaped muon-veto. Figure \ref{fig:deadTime} shows the resulting dead time for a number of detector technologies with different pulse rise times. 
For a cryogenic bolometer of CRESST-II with a mass of 300\,g,  its slow signal implies a dead time above 10\% even for small setups (red line in Figure \ref{fig:deadTime}). 
For the existing \nucleus{} prototype detector, the dead time is predicted to be about 7\% for a cubic muon-veto of 1\,m side length.

The rise time in \nucleus{} detectors can be influenced by changing the phonon collection area of the TES. A larger area leads to a quicker collection of signal phonons and thus a faster pulse rise time. For the \nucleus{} experiment, a rise time of 100~$\mu$s is envisioned. Under these conditions, the dead time, e.g. for a cubic-shaped muon-veto of 1\,m side length, is about 2\% and stays below 10\% even for a dimension of (2\,m)$^3$.  

\section{Sensitivity of \nucleus{} to \cenns{} at the VNS}
\label{sec:Sensitivity}
The \nucleus{} experiment will proceed in a staged approach, as presented in Section~\ref{sec:NucleusDetectors}: a first phase called \nucleus{}-10g, and an upgrade called \nucleus{}-1kg.
At the VNS, with distances of 72~m and 102~m to the two 4.25 GW$_\mathrm{th}$ cores of the CHOOZ nuclear power plant, \nucleus{}-10g will observe 0.33 (0.03) \cenns{}-induced nuclear recoil events per day above 10~eV in the \cawo{} (\alo{}) array.
The high \cenns{} rate in the CaWO$_4$ array, induced by the large tungsten nuclei, allows for a fast discovery of \cenns{} at the VNS. The strongly suppressed \cenns{} rate in the Al$_2$O$_3$ array provides an effective ``neutrino-off'' reference spectrum, which is of great advantage in case of an unknown background shape, as discussed in Section~\ref{sec:SensUnknownBg}.

\subsection{Sensitivity of \nucleus{}-10g for Different Flat Backgrounds}
\label{sec:SensFlatBg}

\begin{figure}
 \centering
 \includegraphics[width=0.75\textwidth]{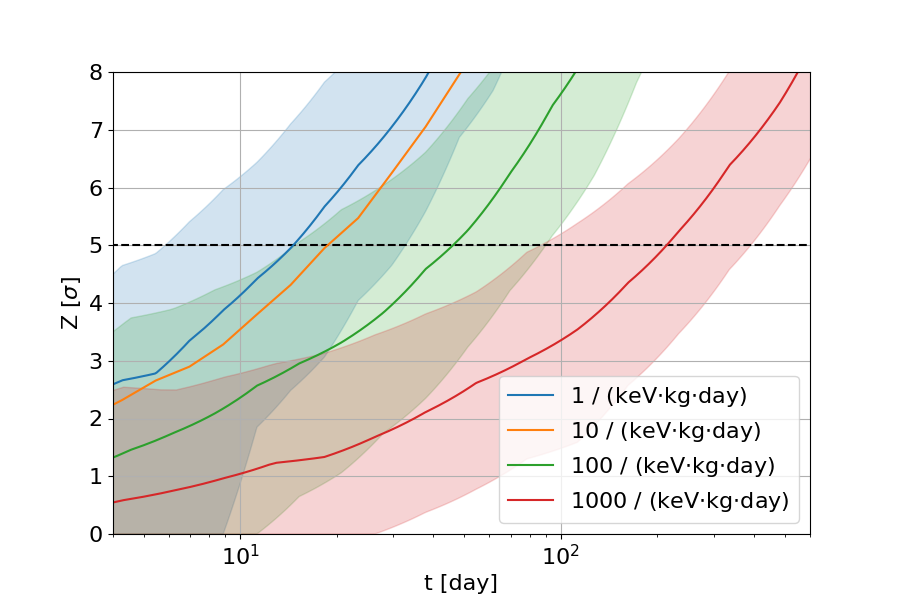}
 \caption{Statistical significance of \cenns{} observation as a function of live time for \nucleus{}-10g, for different background indices, using the expected neutrino flux at the VNS and an energy threshold of 10~eV. For each background index, the median line and 90\% probability bands (computed  from 600 simulated random spectra at each point) are shown. For the background index of 10~\rateunit, only the median line is shown for clarity.}
 \label{fig:stat_signi}
\end{figure}

The results of a sensitivity study for \nucleus{}-10g, similar to the sensitivity study of Reference~\cite{Strauss:2017cuu}, are shown in Figure~\ref{fig:stat_signi} for different background indices.
For each background index and exposure time, 600 random spectra with a flat background and the Standard Model \cenns{} expectation at the VNS are generated for the two target materials \cawo{} and \alo{} in the energy range from 10~eV to 2~keV. To each pair of spectra, a flat-background plus \cenns{} signal strength model (''free model``) and a flat-background-only model (``null model'') are fit. The ``null model'' has only one free parameter, the flat background rate, which is common to both target materials. The ``free model'' has one additional parameter, the signal normalization, for which a value of 1 corresponds to the Standard Model \cenns{} expectation in each target material. The ratio of the maximum likelihood for these nested models ($\mathcal{L}_\mathrm{free}$ and $\mathcal{L}_\mathrm{null}$ respectively) is converted to a significance of \cenns{} observation using $Z = \sqrt{2\cdot\log\left(\mathcal{L}_\mathrm{free} / \mathcal{L}_\mathrm{null}\right)}$. The median, 5$_\mathrm{th}$ and 95$_\mathrm{th}$ percentile of the resulting $Z$ distribution are used to draw the color bands (corresponding to a 90\% probability interval). 

For the most pessimistic background index of 1000~\rateunit, the signal rate in \cawo{} exceeds the background rate only close to the assumed energy threshold of 10~eV, so that long exposures are necessary to distinguish the contributions by spectral shape. Below around 10~\rateunit, the experiment is background-free over most of the region of interest, so that the necessary live time becomes limited by signal statistics.

For a benchmark background rate of 100~\rateunit, a 5$\sigma $ observation of \cenns{} at VNS can be achieved in less than 40 days of measurement time.

\subsection{Achievable Precision of \nucleus{}-10g and -1kg}

As an absolute-rate experiment with the goal of studying the \cenns{} cross-section, \nucleus{} has to rely on reactor neutrino flux predictions which will contribute a significant systematic uncertainty to the measurement. %
At energies above the IBD threshold, the reactor antineutrino spectrum is modeled with an uncertainty of 2-3\%~\cite{Mueller:2011nm,Huber:2011wv}. 
As nuclear recoils on tungsten induced by neutrinos below the IBD threshold extend up to 38~eV in nuclear recoil energy, \nucleus{} has the potential to become sensitive to this as yet unobserved low-energy part of the spectrum. An effort to predict the fluxes of reactor antineutrinos below the IBD threshold is thus necessary to interpret the \nucleus{} data at these lowest recoil energies.

Assuming a flat background of 100~\rateunit{} and an energy threshold of 10~eV, the statistical precision achievable with \nucleus{}-10g after one year is 11\%. Therefore, \nucleus{}-10g is expected to be limited by statistics over its complete measurement time. On the contrary, for \nucleus{}-1kg (with an assumed improved background index of 1~\rateunit{} and an energy threshold of 4~eV) the achievable statistical precision with one year of data is 0.64\%, so that this stage of the experiment will be limited by systematics early on in the measurement campaign. 

\begin{figure}
 \centering
 \includegraphics[width=0.75\textwidth]{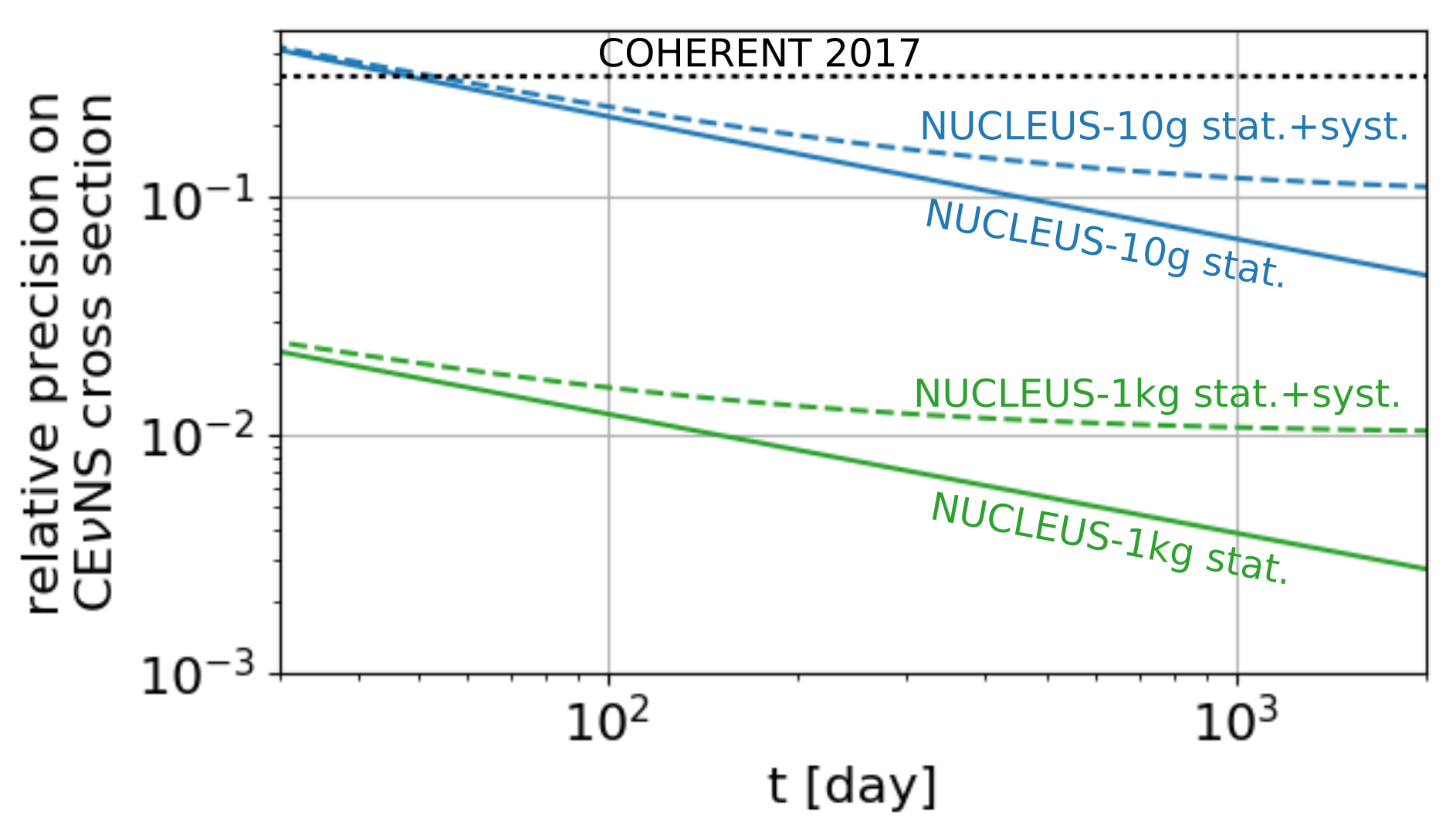}
 \caption{Relative precision of the \cenns{} cross-section measurement as a function of live time for the experimental stages \nucleus-10g (composed of \cawo{} and \alo{} detectors), and \nucleus-1kg (modeled as a Ge detector array). The solid lines show statistical uncertainties (one standard deviation) only, for the dashed lines a systematic uncertainty of 10\% (1\%) is added in case of \nucleus-10g (\nucleus-1kg). The dotted horizontal line indicates the 32\% precision achieved by COHERENT (adding in quadrature the 16\% experimental uncertainty and the 28\% uncertainty of the rate prediction~\cite{Akimov:2017ade}).}
 \label{fig:stat_prec}
\end{figure}

Figure~\ref{fig:stat_prec} shows the precision (one standard deviation) on the \cenns{} cross-section achievable by \nucleus{}, compared to the best current precision achieved by COHERENT~\cite{Akimov:2017ade} (black dashed line). For \nucleus{}-10g (blue), the solid line shows statistical precision only, while the dashed line adds a 10\% systematical uncertainty in quadrature. \nucleus{}-10g can approach the expected systematic limit within few years of live time.
The continuous green line in Figure~\ref{fig:stat_prec} shows the statistical precision achievable with a 1~kg Ge target at the VNS. Adding an optimistic 1\% systematical uncertainty (dashed green line) shows that \nucleus{}-1kg can reach a percent-level measurement of the \cenns{} cross-section within few years of measurement time. The final precision on the \cenns{} cross-section that can be reached by \nucleus{}-1kg depends strongly on the control of systematics, such as neutrino flux normalization, precise knowledge of the energy scale and modeling of backgrounds.

\subsection{\nucleus{}-10g in the Case of a Non-flat Background}
\label{sec:SensUnknownBg}

\begin{figure}
 \centering
 \includegraphics[width=0.75\textwidth]{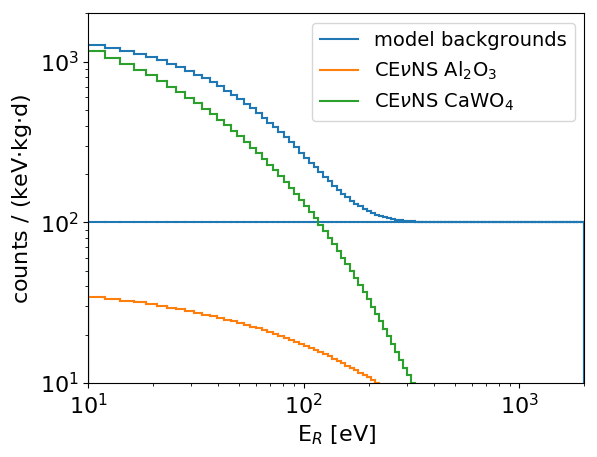}
 \caption{Comparison of \cenns{} recoil spectra and background models discussed in section~\ref{sec:Sensitivity} in the context of \nucleus{}-10g. The flat background of 100~\rateunit{} taken as a benchmark model in section~\ref{sec:SensFlatBg} is well above \cenns{} on \alo{}, while the \cawo{} recoil spectrum rises above it below 110~eV. The exponential background model (see text) with an amplitude of 1500~\rateunit{} and a slope of 45~eV is similar in shape and normalization to the \cawo{} \cenns{} signal, therefore it is adopted for the study of \nucleus{}-10g with a non-flat background in section~\ref{sec:SensUnknownBg} (see text).}
 \label{fig:bg_models}
\end{figure}

Since \nucleus{} is aiming to measure a signal in a previously unexplored energy range in a radiogenically challenging environment, assuming a flat background down to threshold may well be overly optimistic. Accepting a background model with a free shape parameter drastically reduces the detection significance in a single target material, as the signal can then also be fit reasonably well with the background-only model.

The multi-target approach of \nucleus{}-10g (i.e. deploying CaWO$_4$ as well as Al$_2$O$_3$ target arrays) has the potential to mitigate the impact of a background of unknown shape. With a \cenns{} signal more than an order of magnitude below the one in CaWO$_4$ (see Fig.~\ref{fig:RatePlot}), Al$_2$O$_3$ can be conservatively regarded as measuring background under identical experimental conditions to the CaWO$_4$ array. Precise knowledge about the rate and shape of the spectra associated with the various background sources in the two materials is needed to apply the knowledge from this "side-band region" (Al$_2$O$_3$ spectrum) to constraining the fit in the signal region (CaWO$_4$ spectrum). This requires careful simulations and calibration campaigns using both target materials in different environments.

Assuming for simplicity an identical appearance of backgrounds in \alo{} and \cawo{} allows to study the power of this multi-target approach in \nucleus{}-10g. To demonstrate the advantage of combining materials with different expected \cenns{} signal strength, a scenario is considered here in which the background is non-flat and, more importantly, of unknown shape. 

As a simple non-flat background model an exponential + flat model with the functional form $R(E)=C+A\cdot\mathrm{e}^{-E/B}$ is considered, where $C$ is the flat background rate, $A$ is the amplitude and $B$ is the slope of the exponential. Changing the slope of the exponential background while keeping the total integral fixed and the flat background level constant at 100~\rateunit{}, shows the strongest impact on the sensitivity of \nucleus{}-10g after one year to occur for a slope of around 45~eV. Therefore, to be conservative, a background of this slope is chosen for the non-flat scenario. An amplitude of 1500~\rateunit{} is set for the exponential background (see Figure~\ref{fig:bg_models}) so that the background level is higher than the signal rate at all observed energies. In this scenario, there are 120 (636) \cenns{} (background) interactions in the 6~g \cawo{} array expected between 10~eV and 2~keV within one year. In the 4~g \alo{} array, these numbers change to 12 (415) \cenns{} (background) interactions.

The \cawo{} spectrum alone is fit nearly equally well by both the (three parameter) background-only model and the (four parameter) model including the signal. Thus no sensitivity to the \cenns{} cross-section can be derived from \cawo{} alone.
The simultaneous fit of both spectra uses the same background model in both materials (three parameters) and the SM \cenns{} expectation for each material scaled by a common strength parameter. In this way, the \alo{} array helps constraining the background rate and shape, so that an expected preference for the background + \cenns{} model of more than 4~$\sigma$ can be extracted after one year.

This result shows the crucial advantage of the multi-target approach: the sensitivity is significantly improved even without any prior knowledge of the background shape (apart from the flat + exponential - parameterization). Even in case of a background with a signal-like shape and strength, \cenns{} becomes observable through its characteristic dependence on the nuclear composition of the target. This advantage is complementary to the traditional concept of particle identification, i.e. distinction of electron recoil and nuclear recoil events. 
While the multi-target approach is statistical only, it can be used to constrain neutron background which features an identical experimental signature as CEvNS, i.e. a  nuclear recoil, but a distinct dependence on the nuclear composition of different materials.


\section{Conclusion}
\label{sec:Conclusion}
The VNS at the Chooz nuclear plant is a promising new experimental site for the planned \cenns{} experiment \nucleus{}. The site is located in  close distance to the two 4.25\,GW$\mathrm{th}$ reactor cores of the Chooz nuclear power plant in France, providing an antineutrino flux of $\sim~10^{12}\,\overline{\nu}/(\mathrm{s}\,\cdot\,\mathrm{cm}^2 )$.  
First muon attenuation measurements on-site indicate a shallow overburden of about 3\,m.w.e. at the VNS.

The \nucleus{} detector concept provides a suitable technology for a new-generation \cenns{} experiment at the VNS. 
Thanks to its unprecedentedly low energy threshold of $\leq20\,\mathrm{eV}$, a strong \cenns{} signal is expected which allows a significant miniaturization of the detector size.
Dedicated MC simulations show that the muon-induced dead time is expected to stay well below 10\%, given the fast rise-time of the \nucleus{} detectors. The small size of the target detectors and the envisioned compact setup, which consists of passive and active shielding material, greatly reduce the overall size of the experiment matching the space constraints of the VNS. 
Using active and passive background reduction techniques, the \nucleus{} experiment aims to reach a background count rate of $\leq$\,100\,\rateunit{}.

In the first phase of the experiment, \nucleus{}-10g aims at a first observation of \cenns{} at a nuclear reactor with a total target mass of 10\,g. 
Due to the demonstrated threshold in the 10\,eV regime, \nucleus{} will probe the reactor antineutrino spectrum for the first time below 1.8\,MeV - below the threshold of inverse beta-decay. 
The reach of \nucleus{} strongly depends on the achieved background level at low energy. \nucleus{}-10g will explore this background for the first time at energies below 100\,eV and at shallow overburden.
A statistical analysis shows that \cenns{} can be observed within two weeks of live time assuming our benchmark background model. 
Even in a scenario in which shape and rate of the background are similar to that of the \cenns{} signal, \nucleus{}-10g reaches a sensitivity of more than 4$\sigma$ for \cenns{}  after one year of measuring time. 
\nucleus{}-10g, which is expected to take data from 2021 on,  has the potential to achieve a final precision of $\sim$10\% on the \cenns{} cross-section.

The second phase, \nucleus{}-1kg, is planned for commissioning in 2023 and aims 
for precision measurements of the \cenns{} cross-section at a level of 1\%. 
Harnessing the full potential of \nucleus{}-1kg requires a significant reduction of systematic uncertainties. Particularly, to take advantage of the achievable statistical precision, an improved prediction of the reactor anti-neutrino spectrum is needed.
This opens the door for the study of physics beyond the Standard Model of Particle Physics such as the search for non-standard neutrino interactions, a neutrino magnetic dipole moment or exotic neutral currents, as well as a test of the Reactor Antineutrino Anomaly~\cite{Mention:2011rk}. 

\nucleus{} will demonstrate a new detector technology with ultra-low thresholds and allow to study fundamental properties of the neutrino, pushing the low-energy frontier in neutrino observations.

\section{Acknowledgements}

We thank the EdF staff and management board of the Chooz nuclear power plant for their continuous support during the VNS background measurement campaigns. We are also grateful to the Laboratoire de Physique Nucléaire et des Hautes \'Energies (LPNHE, Paris Jussieu) for kindly providing us with the cosmic wheel.

This work has been supported through the DFG by the SFB1258 and the Excellence Cluster Universe and by the BMBF 05A17WO4. The \nucleus{} experiment is funded by the ERC-StG2018-804228 "NUCLEUS".
We acknowledge the financial support for BASKET provided by the Cross-Disciplinary Program on Instrumentation and Detection of CEA, the French Alternative Energies and Atomic Energy Commission. 
V. Wagner was supported by a CEA/Marie-Curie Eurotalents Fellowship.

\bibliographystyle{unsrt}
\bibliography{VNS}

\end{document}